\documentclass{lmcs} %
\pdfoutput=1

\usepackage{lastpage}
\lmcsdoi{20}{1}{2}
\lmcsheading{}{\pageref{LastPage}}{}{}%
{Dec.~07,~2022}{Jan.~11,~2024}{}

\keywords{offline monitoring, online monitoring, energy-aware monitoring, cyber-physical systems, formal methods}
\usepackage[utf8]{inputenc}
\usepackage[english]{babel}

\newcommand{\reach}[1]{\textbf{\texttt{overReach}}(#1)}
\newcommand{\logg}{\ensuremath{\ell}}
\newcommand{\MoULDyS}{\textbf{\texttt{MoULDyS}}} %

\newcommand{\flowstar}{\textbf{\texttt{Flow*}}}

\definecolor{algocommentscolor}{rgb}{.2, .2, 1}

\usepackage[ruled,vlined,linesnumbered]{algorithm2e}
	\SetKwInOut{Input}{input}
	\SetKwInOut{Output}{output}
	
	\SetCommentSty{mycommfont}

\usepackage{subcaption}

\usepackage{float} %

\usepackage{paralist} %

\newenvironment{ienumerate}
	{\begin{inparaenum}[\itshape i\upshape)]}
	{\end{inparaenum}}

\newenvironment{oneenumerate}
	{\begin{inparaenum}[1)]}
	{\end{inparaenum}}

\usepackage{amsmath} %
\usepackage{amssymb} %

\usepackage{graphicx}
\graphicspath{{figures}}

\definecolor{darkblue}{rgb}{0, 0, 0.7}

\usepackage[%
		pdfauthor={Bineet Ghosh and Étienne andre},%
		pdftitle={Offline and online energy-efficient monitoring of scattered uncertain logs using a bounding model},
		breaklinks  = true,
		colorlinks  = true,
		citecolor   = blue!50!blue,
		linkcolor   = darkblue,
		urlcolor    = blue!50!green,
	]{hyperref}

\usepackage[capitalise,english,nameinlink]{cleveref} %
\crefname{line}{\text{line}}{\text{lines}} %
\crefname{defi}{\text{Definition}}{\text{Definitions}} %
\crefname{rem}{\text{Remark}}{\text{Remarks}} %

\usepackage{tikz}
\usetikzlibrary{arrows,automata,calc,positioning} %

\tikzstyle{sample}=[color=blue,radius=7pt]
\tikzstyle{missingsample}=[color=blue!20,radius=4pt]
\tikzstyle{extrasample}=[color=green,radius=4pt]
\tikzstyle{signal}=[color=blue,-,densely dotted]
\tikzstyle{nodraw}=[draw=none,inner sep=0pt,minimum size=0pt]
\tikzstyle{uncertainsample}=[draw=blue,thick] %

\newcommand{\HiLi}[2][yellow]{\begingroup
	\setlength{\fboxsep}{1pt}
	\colorbox{#1}{#2}
	\endgroup}

\theoremstyle{plain} %

	\definecolor{colorok}{RGB}{0,0,0}

\newcommand{\eg}{\textcolor{colorok}{e.g.,}\xspace}
\newcommand{\etal}{\textcolor{colorok}{\emph{et al.}}\xspace}
\newcommand{\ie}{\textcolor{colorok}{i.e.,}\xspace}
\newcommand{\st}{\textcolor{colorok}{s.t.}\xspace}

\theoremstyle{defC}
\newtheorem{exaC}[thm]{Example}

\begin{document}

\title[Monitoring of uncertain logs using a bounding model]{Offline and online energy-efficient monitoring of scattered uncertain logs using a bounding model}
\author[B.~Ghosh]{Bineet Ghosh\lmcsorcid{0000-0002-1371-2803}}[a,b]
\author[É.~André]{Étienne André\lmcsorcid{0000-0001-8473-9555}}[c,d]

\address{The University of Alabama, AL, The United States of America}	%
\email{bineet@ua.edu}  %

\address{The University of North Carolina at Chapel Hill, NC, The United States of America}	

\address{Université de Lorraine, CNRS, Inria, LORIA, F-54000 Nancy, France}	%
\address{Université Sorbonne Paris Nord, LIPN, CNRS UMR 7030, F-93430 Villetaneuse, France}	%
\begin{abstract}
	Monitoring the correctness of distributed cyber-physical systems is essential.
	Detecting possible safety violations can be hard when some samples are uncertain or missing.
	We monitor here black-box cyber-physical system, with logs being uncertain both in the state and timestamp dimensions: that is, not only the logged value is known with some uncertainty, but the time at which the log was made is uncertain too.
	In addition, we make use of an over-approximated yet expressive model, given by a non-linear extension of dynamical systems.
		Given an offline log, our approach is able to monitor the log against safety specifications with a limited number of false alarms.
	As a second contribution, we show that our approach can be used online to minimize the number of sample triggers, with the aim at energetic efficiency.
	We apply our approach to three benchmarks, an anesthesia model, an adaptive cruise controller and an aircraft orbiting system.
\end{abstract}

\maketitle{}

\section{Introduction}\label{section:introduction}

The pervasiveness of distributed cyber-physical systems is highly increasing, accompanied by associated safety concerns.
Formal verification techniques usually require a (white-box) model, which may not often available, because some components are black-box, or because the entire system has no formal model.
In addition, despite some success in verifying formal models from the industry in the recent past (\eg{} \cite{BCMDH92,KGNTWPSTFRN09,LLN18,ACFJL21}), formal verification techniques for cyber-physical systems are often subject to state space explosion, often preventing a satisfactory scalability (see \eg{} \cite{Pelanek08,CKNZ11}).
Therefore, \emph{monitoring}, as a lightweight yet feasible verification technique, can bring practical results of high importance for larger models.

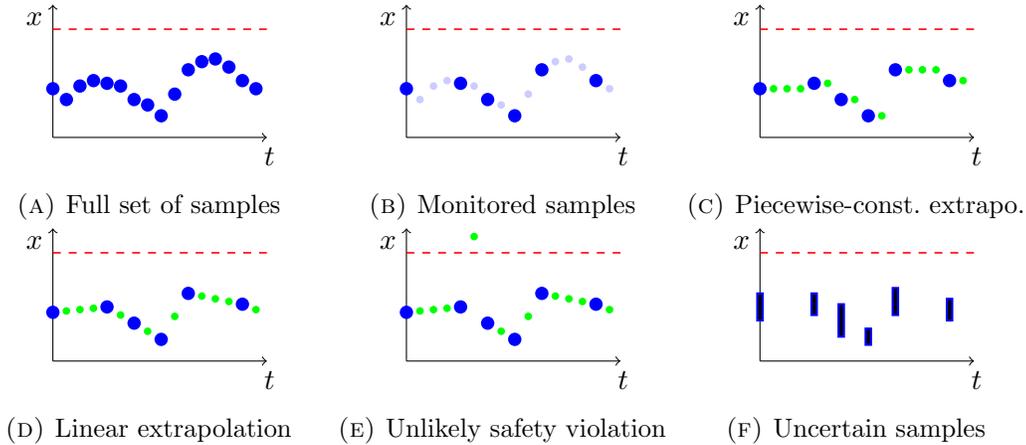
\begin{figure*}[tb]
	\newcommand{\imagewidth}{.3\linewidth}
	\newcommand{\tikzscale}{.6}
	\begin{subfigure}[c]{\imagewidth}
		\centering
		\begin{tikzpicture}[shorten >=1pt, scale=\tikzscale, yscale=.6, xscale=0.6, auto]

			\draw[->] (0, 0) --++ (0, 5.0) node[anchor=north east]{$x$};
			\draw[->] (0, 0) --++ (8.0, 0) node[anchor=north]{$t$};
			\draw[dashed, color=red, semithick] (0, 4) --++ (8.0, 0);

			\fill[sample] (0, 1.8) coordinate (s1) circle[];
			\fill[sample] (0.5, 1.4) coordinate (s2) circle[];
			\fill[sample] (1, 1.9) coordinate (s3) circle[];
			\fill[sample] (1.5, 2.1) coordinate (s4) circle[];
			\fill[sample] (2, 2) coordinate (s5) circle[];
			\fill[sample] (2.5, 1.9) coordinate (s6) circle[];
			\fill[sample] (3, 1.4) coordinate (s7) circle[];
			\fill[sample] (3.5, 1.2) coordinate (s8) circle[];
			\fill[sample] (4, 0.8) coordinate (s9) circle[];
			\fill[sample] (4.5, 1.6) coordinate (s10) circle[];
			\fill[sample] (5, 2.5) coordinate (s11) circle[];
			\fill[sample] (5.5, 2.8) coordinate (s12) circle[];
			\fill[sample] (6.0, 2.9) coordinate (s13) circle[];
			\fill[sample] (6.5, 2.6) coordinate (s14) circle[];
			\fill[sample] (7, 2.1) coordinate (s15) circle[];
			\fill[sample] (7.5, 1.8) coordinate (s16) circle[];

		\end{tikzpicture}

		\caption{Full set of samples}
		\label{example:full}
	\end{subfigure}
	\begin{subfigure}[c]{\imagewidth}
		\centering
		\begin{tikzpicture}[shorten >=1pt, scale=\tikzscale, yscale=.6, xscale=0.6, auto]

			\draw[->] (0, 0) --++ (0, 5.0) node[anchor=north east]{$x$};
			\draw[->] (0, 0) --++ (8.0, 0) node[anchor=north]{$t$};
			\draw[dashed, color=red, semithick] (0, 4) --++ (8.0, 0);

			\fill[sample] (0, 1.8) coordinate (s1) circle[];
			\fill[missingsample] (0.5, 1.4) coordinate (s2) circle[];
			\fill[missingsample] (1, 1.9) coordinate (s3) circle[];
			\fill[missingsample] (1.5, 2.1) coordinate (s4) circle[];
			\fill[sample] (2, 2) coordinate (s5) circle[];
			\fill[missingsample] (2.5, 1.9) coordinate (s6) circle[];
			\fill[sample] (3, 1.4) coordinate (s7) circle[];
			\fill[missingsample] (3.5, 1.2) coordinate (s8) circle[];
			\fill[sample] (4, 0.8) coordinate (s9) circle[];
			\fill[missingsample] (4.5, 1.6) coordinate (s10) circle[];
			\fill[sample] (5, 2.5) coordinate (s11) circle[];
			\fill[missingsample] (5.5, 2.8) coordinate (s12) circle[];
			\fill[missingsample] (6.0, 2.9) coordinate (s13) circle[];
			\fill[missingsample] (6.5, 2.6) coordinate (s14) circle[];
			\fill[sample] (7, 2.1) coordinate (s15) circle[];
			\fill[missingsample] (7.5, 1.8) coordinate (s16) circle[];

		\end{tikzpicture}

		\caption{Monitored samples}
		\label{example:monitored}
	\end{subfigure}
	\begin{subfigure}[c]{\imagewidth}
		\centering
		\begin{tikzpicture}[shorten >=1pt, scale=\tikzscale, yscale=.6, xscale=0.6, auto]

			\draw[->] (0, 0) --++ (0, 5.0) node[anchor=north east]{$x$};
			\draw[->] (0, 0) --++ (8.0, 0) node[anchor=north]{$t$};
			\draw[dashed, color=red, semithick] (0, 4) --++ (8.0, 0);

			\fill[sample] (0, 1.8) coordinate (s1) circle[];
			\fill[extrasample] (0.5, 1.8) coordinate (s2) circle[];
			\fill[extrasample] (1, 1.8) coordinate (s3) circle[];
			\fill[extrasample] (1.5, 1.8) coordinate (s4) circle[];
			\fill[sample] (2, 2) coordinate (s5) circle[];
			\fill[extrasample] (2.5, 2) coordinate (s6) circle[];
			\fill[sample] (3, 1.4) coordinate (s7) circle[];
			\fill[extrasample] (3.5, 1.4) coordinate (s8) circle[];
			\fill[sample] (4, 0.8) coordinate (s9) circle[];
			\fill[extrasample] (4.5, 0.8) coordinate (s10) circle[];
			\fill[sample] (5, 2.5) coordinate (s11) circle[];
			\fill[extrasample] (5.5, 2.5) coordinate (s12) circle[];
			\fill[extrasample] (6.0, 2.5) coordinate (s13) circle[];
			\fill[extrasample] (6.5, 2.5) coordinate (s14) circle[];
			\fill[sample] (7, 2.1) coordinate (s15) circle[];
			\fill[extrasample] (7.5, 2.1) coordinate (s16) circle[];

		\end{tikzpicture}

		\caption{Piecewise-const.\ extrapo.}
		\label{example:linear}
	\end{subfigure}

	\begin{subfigure}[c]{\imagewidth}
		\centering
		\begin{tikzpicture}[shorten >=1pt, scale=\tikzscale, yscale=.6, xscale=0.6, auto]

			\draw[->] (0, 0) --++ (0, 5.0) node[anchor=north east]{$x$};
			\draw[->] (0, 0) --++ (8.0, 0) node[anchor=north]{$t$};
			\draw[dashed, color=red, semithick] (0, 4) --++ (8.0, 0);

			\fill[sample] (0, 1.8) coordinate (s1) circle[];
			\fill[extrasample] (0.5, 1.85) coordinate (s2) circle[];
			\fill[extrasample] (1, 1.9) coordinate (s3) circle[];
			\fill[extrasample] (1.5, 1.95) coordinate (s4) circle[];
			\fill[sample] (2, 2) coordinate (s5) circle[];
			\fill[extrasample] (2.5, 1.7) coordinate (s6) circle[];
			\fill[sample] (3, 1.4) coordinate (s7) circle[];
			\fill[extrasample] (3.5, 1.1) coordinate (s8) circle[];
			\fill[sample] (4, 0.8) coordinate (s9) circle[];
			\fill[extrasample] (4.5, 3.3/2) coordinate (s10) circle[];
			\fill[sample] (5, 2.5) coordinate (s11) circle[];
			\fill[extrasample] (5.5, 2.4) coordinate (s12) circle[];
			\fill[extrasample] (6.0, 2.3) coordinate (s13) circle[];
			\fill[extrasample] (6.5, 2.2) coordinate (s14) circle[];
			\fill[sample] (7, 2.1) coordinate (s15) circle[];
			\fill[extrasample] (7.5, 1.9) coordinate (s16) circle[];

		\end{tikzpicture}

		\caption{Linear extrapolation}
		\label{example:constant}
	\end{subfigure}
	\begin{subfigure}[c]{\imagewidth}
		\centering
		\begin{tikzpicture}[shorten >=1pt, scale=\tikzscale, yscale=.6, xscale=0.6, auto]

			\draw[->] (0, 0) --++ (0, 5.0) node[anchor=north east]{$x$};
			\draw[->] (0, 0) --++ (8.0, 0) node[anchor=north]{$t$};
			\draw[dashed, color=red, semithick] (0, 4) --++ (8.0, 0);

			\fill[sample] (0, 1.8) coordinate (s1) circle[];
			\fill[extrasample] (0.5, 1.85) coordinate (s2) circle[];
			\fill[extrasample] (1, 1.9) coordinate (s3) circle[];
			\fill[extrasample] (1.5, 1.95) coordinate (s4) circle[];
			\fill[sample] (2, 2) coordinate (s5) circle[];
			\fill[extrasample] (2.5, 4.6) coordinate (s6) circle[];
			\fill[sample] (3, 1.4) coordinate (s7) circle[];
			\fill[extrasample] (3.5, 1.1) coordinate (s8) circle[];
			\fill[sample] (4, 0.8) coordinate (s9) circle[];
			\fill[extrasample] (4.5, 3.3/2) coordinate (s10) circle[];
			\fill[sample] (5, 2.5) coordinate (s11) circle[];
			\fill[extrasample] (5.5, 2.4) coordinate (s12) circle[];
			\fill[extrasample] (6.0, 2.3) coordinate (s13) circle[];
			\fill[extrasample] (6.5, 2.2) coordinate (s14) circle[];
			\fill[sample] (7, 2.1) coordinate (s15) circle[];
			\fill[extrasample] (7.5, 1.9) coordinate (s16) circle[];

		\end{tikzpicture}

		\caption{Unlikely safety violation}
		\label{example:unsafe}
	\end{subfigure}
	\begin{subfigure}[c]{\imagewidth}
		\centering
		\begin{tikzpicture}[shorten >=1pt, scale=\tikzscale, yscale=.6, xscale=0.6, auto]

			\draw[->] (0, 0) --++ (0, 5.0) node[anchor=north east]{$x$};
			\draw[->] (0, 0) --++ (8.0, 0) node[anchor=north]{$t$};
			\draw[dashed, color=red, semithick] (0, 4) --++ (8.0, 0);

			\newcommand{\uncertainsample}[3]{\fill[uncertainsample] (#1-.1, #2-#3) rectangle (#1+.1, #2+#3);}
			
			\uncertainsample{0}{2}{0.5};
			\uncertainsample{2}{2.1}{0.4};
			\uncertainsample{3}{1.5}{0.6};
			\uncertainsample{4}{0.9}{0.3};
			\uncertainsample{5}{2.2}{0.5};
			\uncertainsample{7}{1.9}{0.4};
		\end{tikzpicture}

		\caption{Uncertain samples}
		\label{example:uncertain}
	\end{subfigure}
	\caption{Monitoring at discrete time steps}
	\label{figure:monitoring-example}
\end{figure*}

Monitoring aims at analyzing the log of a concrete system, so as to deduce whether a specification (\eg{} a safety property) is violated~\cite{BDDFMNS18}.
Monitoring can be done \emph{offline} (\ie{} after the system execution, assuming the knowledge of the entire log, see \eg{} \cite{BCEHKM16}), or \emph{online} (at runtime, assuming a partial log); see \cite{Maler16} for a discussion on online verification.
When the log is an aperiodic timed sequence of valuations of continuous variables, with a logging \emph{not} occurring at every discrete time step, and when the system under monitoring is a black box, a major issue is: how to be certain that, in between two discrete valuations, the specification was not violated at another discrete time step at which no logging was performed?
For example, consider a system for which a logging at every discrete time step would yield the log depicted in \cref{example:full}.
Assume the logging was done at only \emph{some} time steps, given in \cref{example:monitored}, due to some sensor faults, or to save energy with only a sparse, scattered logging.
How to be certain that, in between two discrete samples, another discrete sample (not recorded) did not violate the specification?
For example, by just looking at the discrete samples in \cref{example:monitored}, there is no way to formally guarantee that the unsafe zone (\ie{} above the red, dashed line) was never reached by another discrete sample which was not recorded.
In many practical cases, a piecewise-constant or linear approximation (see, \eg{} \cref{example:linear,example:constant}, where the large blue dots denote actual samples, while the small green dots denote reconstructed samples using some extrapolation) is arbitrary and not appropriate; even worse, it can yield a ``safe'' answer, while the actual system could actually have been unsafe at some of the missing time steps.
On the contrary, assuming a completely arbitrary dynamics will always yield ``potentially unsafe''---thus removing the interest of monitoring.
For example, from the samples in \cref{example:monitored}, without any knowledge of the model, one can always envision the situation in \cref{example:unsafe}, which shows the variable~$x$ crossing the unsafe region (dashed) at some unlogged discrete time step---even though this is unlikely if the dynamics is known to vary ``not very fast''.

\paragraph{Contributions}
In this work, we address the problem of performing monitoring over a set of scattered and \emph{uncertain} samples.
First, we cope with uncertainties from the sensors by allowing for \emph{uncertain} samples, given by zonotopes over the continuous variables; that is, at each logged timestamp, the log gives not a constant value for the continuous variables, but a \emph{zonotope}.\footnote{%
	 A zonotope is a special form of a convex polyhedron that is centrally symmetric.
}
For example, let us examine the case of an adaptive cruise control (ACC).
In ACC, it is crucial to accurately measure the distance to the vehicle in front for maintaining safety.
Nonetheless, due to uncertainties in the sensors, it may not always be feasible to obtain the precise distance of the lead vehicle.
However, it may still be possible to measure this distance within a certain margin of error.
Consequently, the measured distance of the lead vehicle becomes an interval of values that more accurately represents the true distance, taking into account the potential inaccuracies inherent in the sensor's error range.
For instance, when the reading is taken using a sensor with a $\pm 5\%$ error and the value read by the sensor is~$1$, the uncertain value can be calculated by accounting for the sensor's error tolerance, resulting in a range of $[0.95, 1.05]$.
In this work, to represent all state variable values along with their uncertainties, we choose zonotopes to represent our samples.
A simple case of an uncertain log over a single variable~$x$ is depicted in \cref{example:uncertain} in the form of simple intervals.
The uncertainty can come from the error margin of a sensor: even though the read value is constant, one may need to turn it into an interval, when the sensor only guarantees a limited precision.
In addition, the timestamp at each discrete sample of the log can itself be uncertain, in the form of an \emph{interval} (not shown in \cref{example:uncertain}, where the timestamp is punctual).
This second form of uncertainty can come from network latency, or clocks with limited precision.

Second, %
to over-approximate the system behavior, and in the spirit of the ``model-bounded monitoring'' proposed in~\cite{WAH22TCPS},
	we use an extension of \emph{linear dynamical systems}, extended with uncertainty, \ie{} allowing \emph{uncertainty} in the dynamics matrix~\cite{lal}.
Having some over-approximated knowledge of the system is a natural assumption in practice: when monitoring a car, one generally knows an upper-bound on its maximum speed, or on its maximum acceleration (perhaps depending on its current speed).
To cope with the liberal dynamics of our extension of linear dynamical systems, we use a recent technique~\cite{ghosh1}, that performs an efficient %
	reachability analysis for such uncertain linear dynamical systems.
The use of such an over-approximation of the actual system is the crux of our approach, allowing us to discard unlikely behaviors, such as the unlikely safety violation depicted in \cref{example:unsafe}.

Our first main contribution is to propose a new rigorous analysis technique for offline monitoring of safety properties over scattered \emph{uncertain} samples, using uncertain linear systems as an over-approximation of the system.
This over-approximation allows us to extrapolate the behavior since the latest known sample, and to rule out safety violations at some missing discrete samples.
Note that our approach uses some discrete analysis as underlying reachability computation technique, and will not however guarantee the absence of safety violations at arbitrary (continuous) timestamps; its main advantage is to offer formal guarantees in the context of missing discrete samples for a given logging granularity.

Our second main contribution focuses on \emph{energy-efficient online monitoring}.
For each recorded sample, we run a reachability analysis%
, and we derive the smallest next discrete time step~$t$ in the future at which the safety property may be violated depending on the latest known sample and the over-approximated model dynamics.
In a context in which monitoring simply observes the behavior and does not lead to corrective actions, any sample before~$t$ is useless because we \emph{know} from the over-approximated model dynamics that no safety violation can happen before~$t$.
Therefore, we can schedule the next sample at time~$t$, which reduces the number of discrete samples, and therefore the energy consumption and bandwidth use.
We show that our method is correct, \ie{} we can safely discard discrete samples without missing any unsafe behavior.

Our third contribution is the implementation of our algorithms into an original tool \MoULDyS{}~\cite{GHOSH2023102976}.
We then show the practical applicability of our approach on three benchmarks: an anesthesia model, an adaptive cruise controller, and an aircraft orbiting system.
We conduct various experiments to showcase the effectiveness and scalability of our approach across multiple factors. Specifically, these experiments highlight how uncertainty in log samples, uncertainty in timestamps, and the number of log samples affect the performance of our algorithms. This encompasses both the scalability of the algorithms and their ability to accurately verify the correctness of the systems behavior.

\paragraph{About this manuscript}
This manuscript is an extension of~\cite{GA22}.
In addition to several details, we significantly increased the content in two main directions.
\begin{oneenumerate}
	\item We enhance the uncertainty by considering not only uncertainty over the sample valuations (as in~\cite{GA22}), but also over the sample timestamps: in~\cite{GA22}, the timestamp at each discrete sample of the log was supposed to be constant (\ie{} a single point).
	Here, we extend this notion to an \emph{interval}, making our work able to address a \emph{bi-dimensional uncertainty}.
	\item We consider an additional case study of an aircraft orbiting system (new \cref{subsection:aircraft}), to which we notably apply our offline algorithm extended with uncertainty over the timestamps.
	\item We redid all experiments from~\cite{GA22} to remove their randomness: in short, in~\cite{GA22}, our tool was first generating a random log, and then applying our monitoring algorithms to this log.
		We decoupled this aspect in the newer version of \MoULDyS{}~\cite{GHOSH2023102976}, and we generated random logs once for all, and then our tool applies monitoring on these statically generated logs---this allows for exact reproducibility of our results.
\end{oneenumerate}

\paragraph{Outline}
We review related works in \cref{section:related}.
We recall uncertain linear dynamical systems in \cref{sec:prelims}.
We introduce our offline and online monitoring frameworks in \cref{sec:monitoring}, and run experiments in \cref{sec:case_studies}.
We draw perspectives in \cref{section:conclusion}.

\section{Related works}\label{section:related}

\paragraph{Monitoring}
Monitoring complex systems, and notably cyber-physical systems, drew a lot of attention in the last decades, \eg{} \cite{MN04,BKZ17,BDDFMNS18,WAH22TCPS,MCW21}.
	While the main drawback of monitoring is a lack of formal guarantees on the global behavior of a system, its advantage is a much more scalable efficiency compared to techniques such as model checking (see, \eg{} \cite{FBCI20}).
	In addition, monitoring can be performed on black-box systems, the source code (and therefore a model) of which is unavailable.
In parallel to monitoring specifications using signal temporal logics (see \eg{}~\cite{DFM13,JBGNN18,QD20}),
monitoring using automata-based specifications drew recent attention.
Complex, quantitative extensions of automata were studied in the recent years:
		after timed pattern matching on timed regular expressions~\cite{UFAM14} was proposed by Ulus \etal{},
		Waga \etal{} proposed a technique for timed pattern matching~{\cite{WAH16,WHS17,WHS18,Waga19}} (with an additional work by Bakhirkin \etal{}~\cite{BFNMA18}) and then for parametric timed pattern matching~\cite{WA19,WAH23ToSEM}, with application to offline monitoring.
		Then, techniques for pattern matching were lifted to monitoring against complex specification making use of timing parameters and data parameters~\cite{WAH19}.

Monitoring cyber-physical systems also shares some similarities (using different techniques and goals) with conformance testing cyber-physical systems (\eg{} \cite{Dang11,DMP17,ACMMS18}).

In~\cite{WAH22TCPS}, we proposed \emph{model-bounded monitoring}: instead of monitoring a black-box system against a sole specification, we use in addition a (limited, over-approximated) knowledge of the system, to eliminate false positives.
This over-approximated knowledge is given in~\cite{WAH22TCPS} in the form of a \emph{linear hybrid automaton} (LHA)~\cite{HPR94}, an extension of finite-state automata with continuous variables; their flow in each location (``mode'') is given as a linear constraint over derivatives; location invariants and transition guards are given by linear constraints over the system variables.
We use in~\cite{WAH22TCPS} both an \emph{ad-hoc} implementation, and another one based on PHAVerLite~{\cite{Frehse08,BZ19}}.
In this work, we share with~\cite{WAH22TCPS} the principle of using an over-approximation of the model to rule out some violation of the specification.
However, we consider here a different formalism, and we work on discrete samples.
In terms of expressiveness of the over-approximated model:
\begin{ienumerate}%
	\item our approach can be seen as less expressive than~\cite{WAH22TCPS}, in the sense that we have a single (uncertain) dynamics, as opposed to LHAs, where a different dynamics can be defined in each mode; this also allows us to propose a simpler (therefore more efficient) analysis, as each new sample allows us to restart from an exact basis, while in~\cite{WAH22TCPS} at each new sample, the system (from an algorithmic point of view) can be in ``different modes at the same time'';
	\item conversely, our dynamics is also significantly more expressive than the LHA dynamics of~\cite{WAH22TCPS}; we consider not only the class of linear dynamical systems, but even fit into a special case of non-linear systems, by allowing \emph{uncertainty} in the model dynamics---this is what makes our model an over-approximation of the actual behavior.
\end{ienumerate}%
In addition, we also allow for \emph{uncertain} logs in two dimensions:
\begin{oneenumerate}%
	\item uncertain \emph{values}---coping with sensor uncertainties, and
	\item uncertain \emph{timestamps}---coping with local clock uncertainties and/or network delays.
\end{oneenumerate}%
None of these notions of uncertainty were considered in~\cite{WAH22TCPS}.
We also propose a new \emph{ad-hoc} implementation based on~\cite{ghosh1}.

In~\cite{MP16,MP18}, a monitor is constructed from a system model in differential dynamic logic~\cite{Platzer12}.
The main difference between~\cite{MP16,MP18} and our approach relies in the system model: in~\cite{MP16,MP18}, the compliance between the model and the behavior is checked at runtime, while our model is assumed to be an over-approximation of the behavior---which is by assumption compliant with the model.

In~\cite{SWS21}, black-box checking---combining active automata learning and model\linebreak[4]checking---is improved with specification \emph{strengthening}, increasing the chances to obtain an input violating the specification.

\paragraph{Reachability in linear dynamical systems}
In~\cite{ALK11}, given a continuous time linear system with input, the system is discretized and reachable sets for consecutive time intervals are computed. At each step, the \emph{state transition matrix} is expressed using the \emph{Peano-Baker} series. The series is then numerically approximated iteratively using \emph{Riemann sums}. Then a zonotope-based convex hull is computed over-approximating the result of all possible matrices in the uncertain matrix.
In \cite{COMBASTEL20114525}, Combastel and Raka extend an existing algorithm based on zonotopes so that it can efficiently propagate structured parametric uncertainties. As a result, they provide an algorithm for computation of envelopes enclosing the possible states and/or outputs of a class of uncertain linear dynamical systems.
In \cite{lal}, given an uncertain linear dynamical system $\dot{x}=\Lambda_u x$, Lal \etal{} provide a sampling interval $\delta>0$, given an $\epsilon>0$, \st{} the piecewise bilinear function, approximating the solution by interpolating at these sample values, is within $\epsilon$ of the original trajectory.
\cite{ghoshrobustreachset} identifies a class of uncertainties by a set of sufficient conditions on the structure of the dynamics matrix $\Lambda_u$. 
For such classes of uncertainties, the exact reachable set of the linear dynamical system can be computed very efficiently.
But this method is not applicable for arbitrary classes of uncertainties.
In \cite{ghosh1}, given an uncertain linear dynamical system, we provide two algorithms to compute reachable sets. The first method is based on perturbation theory, and the second method leverages a property of linear systems with inputs by representing them as Minkowski sums.
In \cite{ghosh2}, given an uncertain linear dynamical system, we provide an algorithm to compute statistically correct over-approximate reachable sets using \emph{Jeffries Bayes Factor}.
Note that uncertain linear dynamical systems are a special subset of non-linear systems. Thus, uncertain linear dynamical systems can also be modeled as a non-linear system.
Some additional works that deal with computing reachable sets of non-linear systems are \cite{CAS13,TD13,6987596,10.1007/978-3-662-46681-0_5,ARCH15:An_Introduction_to_CORA,KGCC15,CS16}.

\section{Preliminaries}\label{sec:prelims}
In this section, we layout the notations and definitions used in the rest of the paper.
Formal analysis of safety critical systems requires a precise mathematical model of the system, such as linear dynamical systems. But in reality, the precise, exact model is almost never available---parameter variations, sensor and measurement errors, unaccounted parameters are few such causes that make the availability of a precise model impossible.
Presence of such uncertainties in the model makes the safety analysis of these systems useless using traditional methods. Thus, for the analysis to be indeed useful, the safety analysis must consider all possible uncertainties. In~\cite{lal}%
, the authors provide a model, known as \emph{uncertain linear dynamical systems}, to capture such uncertainties.
Consider the following example of an uncertain linear dynamical system.
\begin{exaC}[{\cite[Example~1.1]{ghoshrobustreachset}}]
\label{eg:uls}
Let a discrete linear dynamical system $x^{+} = \Lambda x$, where $\Lambda = \bigl[
\begin{smallmatrix}
    1  & \alpha \\
    0   & 2
\end{smallmatrix}
\bigr]
$
and $\alpha$ represents either the modeling uncertainty or a parameter, assuming $2 \le \alpha \le 3$.
Note that any safety analysis assuming a \emph{fixed} value of $\alpha$ will render the analysis useless---for the safety analysis to be indeed sound, it must consider \emph{all} possible values of $\alpha$, and they cannot be enumerated.
\end{exaC}

Intuitively, uncertain linear dynamical systems model the uncertainties in the system by representing all possible dynamics matrices of the system---clearly, this forms a special class of non-linear dynamical systems. To perform safety analysis of uncertain linear dynamical systems, these works provide reachable set computation techniques that account for all possible uncertainties.

\begin{defi}[Uncertain linear dynamical systems ({\cite[Definition~2.4]{ghoshrobustreachset}})]
\label{def:uls}
An \emph{uncertain linear dynamical system} is denoted as
\begin{equation}
\label{eq:uls}
    x^{+} = \Lambda x
\end{equation}
where $\Lambda \subset \mathbb{R}^{n \times n}$ is the uncertain dynamics matrix.
\end{defi}

\begin{defi}[Reachable set of an uncertain linear dynamical systems (%
	{\cite[Definitions~2.3 and~2.4]{ghoshrobustreachset}}%
)]
Given an initial set $\theta_0$ and time step $t \in \mathbb{Z}$, the reachable set of an uncertain linear dynamical system is defined as:
\begin{eqnarray}
RS(\Lambda, \theta_0, t) = \theta_t =\{ \theta \mid \theta =\xi_{A}(\theta_0, t), A \in \Lambda \}.
\label{eq:reachSet}
\end{eqnarray}
\noindent{}where $\xi_A(\theta_0,t)=A^t \theta_0$.
An alternative definition is:

\begin{equation}
RS(\Lambda, \theta_0, t) = \theta_t = \bigcup_{A \in \Lambda} \xi_A(\theta_0,t).
\label{eq:reachSet2}
\end{equation}

\end{defi}

Note that uncertain linear dynamical systems are capable of modeling systems with parameters or when the system dynamics is not perfectly known---the system has modeling uncertainties.
\cite{lal,ghoshrobustreachset,ghosh1,ghosh2} propose various algorithms to compute reachable sets of these systems that account for uncertainties. In this work, we leverage a recently proposed reachable set computation technique, given in~\cite{ghosh1}, to propose our offline and online monitoring algorithm, primarily due to its efficiency vis-à-vis our setting.

Given an initial set $\theta_0 \subset \mathbb{R}^n$ and given a time step~$t$, we denote by $\theta_t \subset \mathbb{R}^n$ the reachable set of the system (given by \cref{eq:uls}) at time step~$t$.
Next, we define a log of the system with uncertainties in both in system states and timestamps.

\begin{defi}[Uncertain log]\label{definition:log}
Given an uncertain linear dynamical system as in \cref{eq:uls}, a finite length \emph{uncertain log} is defined as follows:
 \[ \logg =\Big\{\big(\hat{\theta}_t,[t^{lb},t^{ub}]\big) \mid \theta_t \subseteq \hat{\theta}_t, \text{for some } t \in [t^{lb},t^{ub}], t^{lb} \le t^{ub} \le H\Big\}\text{,}\]
\noindent{}where $H$ is a given time bound.
\end{defi}

Each tuple $\big(\hat{\theta}_t,[t^{lb},t^{ub}]\big)$ is called a \emph{sample}.
Observe that, both the system state $\hat{\theta}_t$ and timestamp $[t^{lb},t^{ub}]$, in a sample are not necessarily reduced to a \emph{point}.
The length of log $\logg$---number of samples in $\logg$---is given by $|\logg|$.
When considering an uncertain log $\logg$, the $k$-th sample from $\logg$, where $1 \leq k \leq |\logg|$\label{newtext:k-range}, can be represented as $\logg_k = \left(\hat{\theta}_{t_k}, [t_k^{lb}, t_k^{ub}]\right)$. Here, $\hat{\theta}_{t_k}$ denotes an over-approximation of the system state at a specific time step $t_k$, where $t_k$ lies within the interval $[t_k^{lb}, t_k^{ub}]$.
Such a sample denotes that over-approximate state of the system was observed to be $\hat{\theta}_{t_k}$, for some time step $t_k$, where $t_k \in [t_k^{lb},t_k^{ub}]$ (but the exact $t_k$ is not known). This formalism facilitates modeling of situations when the samples are collected over an uncertain channel (such as a shared network with delays) and the precise timestamp is unknown.
Note that the log can be \emph{scattered}---it does not necessarily contain a sample for each $t \in \{1, \dots, H\}$, \ie{} 
$$
\{1, \cdots, H\} \not\subseteq \bigcup_k [t_k^{lb},t_k^{ub}].
$$
We further note that the uncertainties in the logs, arising from the sensor uncertainties of the logging system, are independent of the uncertainties in the system modeling (\cref{def:uls}).
Given two consecutive samples $\logg_k=\big(\hat{\theta}_{t_k},[t_k^{lb},t_k^{ub}]\big)$ and $\logg_{k+1}=\big(\hat{\theta}_{t_{k+1}},[t_{k+1}^{lb},t_{k+1}^{ub}]\big)$, we assume that their timestamps do not intersect, \ie{} $[t_k^{lb},t_k^{ub}] \cap  [t_{k+1}^{lb},t_{k+1}^{ub}] = \emptyset$; or, put differently, $t_k^{ub} < t_{k+1}^{lb}$.

When the samples are being collected locally, timestamps can be precisely known (for all practical purposes).
In such cases---while we still can have uncertainties in the system state---the timestamp can be known precisely.
This leads to a simpler case of logs as defined as follows.

\begin{defi}[Fixed timestamp uncertain log]\label{definition:fixed-log}
A finite length \emph{uncertain log with fixed timestamps} is defined as follows:
 $   \logg =\{(\hat{\theta}_t,t) \mid \theta_t \subseteq \hat{\theta}_t, t \le H\}$,
\noindent{}where $H$ is a given time bound.
\end{defi}

We call a fixed timestamp uncertain log $\logg$ \emph{accurate} if it satisfies the following condition:
$    \forall 1 \le k \le |\logg| : \hat{\theta}_{t_k} = \theta_{t_k}$.
Given an uncertain linear dynamical system, $x^+ = \Lambda x$ with an initial set $\theta_0 \subset \mathbb{R}^{n}$,
	an \emph{over-approximate reachable set of~$x^+$ at time step~$t$} is $\reach{\Lambda,\theta_0, t}$, such that $\theta_t \subseteq \reach{\Lambda,\theta_0, t}$.

\label{newtext:ghosh1}%
In this work, we use the technique proposed in~\cite{ghosh1} to compute $\reach{\Lambda,\theta_0, t}$.
The algorithm from~\cite{ghosh1} first computes the reachable set of the nominal dynamics (which excludes uncertainties), and then computes the reachable set related to the uncertainties in the dynamics. These two sets are then combined using the Minkowski sum to obtain the reachable set of the entire dynamics.
Although computing the reachable set of the nominal dynamics is straightforward, the reachable set related to uncertainties is challenging to compute.
The technique proposed in~\cite{ghosh1} is sound and demonstrates good scalability, as reported in~\cite{ghosh1}, and confirmed by the experiments conducted in our study (see \cref{sec:case_studies}).
Finally note that any technique can be employed to compute reachable sets of uncertain linear systems as long as it is sound.
In other words, as long as the utilized technique is sound, our proposed algorithms remain sound as well.

\paragraph{Safety properties}\label{newtext:safety}
In this work, we are concerned with \emph{safety} properties.
While in many practical cases, a simple threshold over a single variable (or a set of variables) is enough, as in \eg{} \cref{figure:monitoring-example}, we propose a more expressive definition:
a safety property is defined as a \emph{zonotope} over the system variables.
Since our reachable sets are encoded using zonotopes, safety verification will consist in checking intersection over zonotopes.

\section{Monitoring using uncertain linear dynamical systems as bounding model}\label{sec:monitoring} %

In this section, we propose the two main contributions of this work:
\begin{oneenumerate}
    \item \emph{Offline monitoring}: Given an uncertain log{---arising, \eg{} due to faulty sensors and collected over a shared network---}we propose an algorithm to infer the safety of a system as given in \cref{eq:uls}.
    We prove our method's soundness.
    \item \emph{Online monitoring}: We propose a framework to infer safety of a system, as in \cref{eq:uls}, that triggers the logging system to sample only when needed.
    Note that, as we only consider the system at \emph{discrete} time steps, the method cannot be sound nor complete, \ie{} there always exists a small possibility that the system might violate the safety specification in between two concrete samples (this will be discussed in \cref{ss:future}).
    However, our online method \emph{is} both \textit{sound} and \textit{complete} \emph{at the discrete timestamps}, and under the assumption that the samples are free from uncertainties.
    That is, our method infers the system to be \textit{safe} if and only if the actual behavior of the system is safe at any discrete timestamp, when the logging system can generate accurate samples of the system.
    Put it differently, we guarantee that skipping some logging in the future using our method will not remove any sample where a violation could have been observed.
\end{oneenumerate}

\subsection{Offline monitoring over fixed timestamp uncertain logs}\label{ss:offline}

\begin{figure*}[tb]
    \begin{subfigure}{.65\textwidth}
        \centering

        \includegraphics[width=\linewidth]{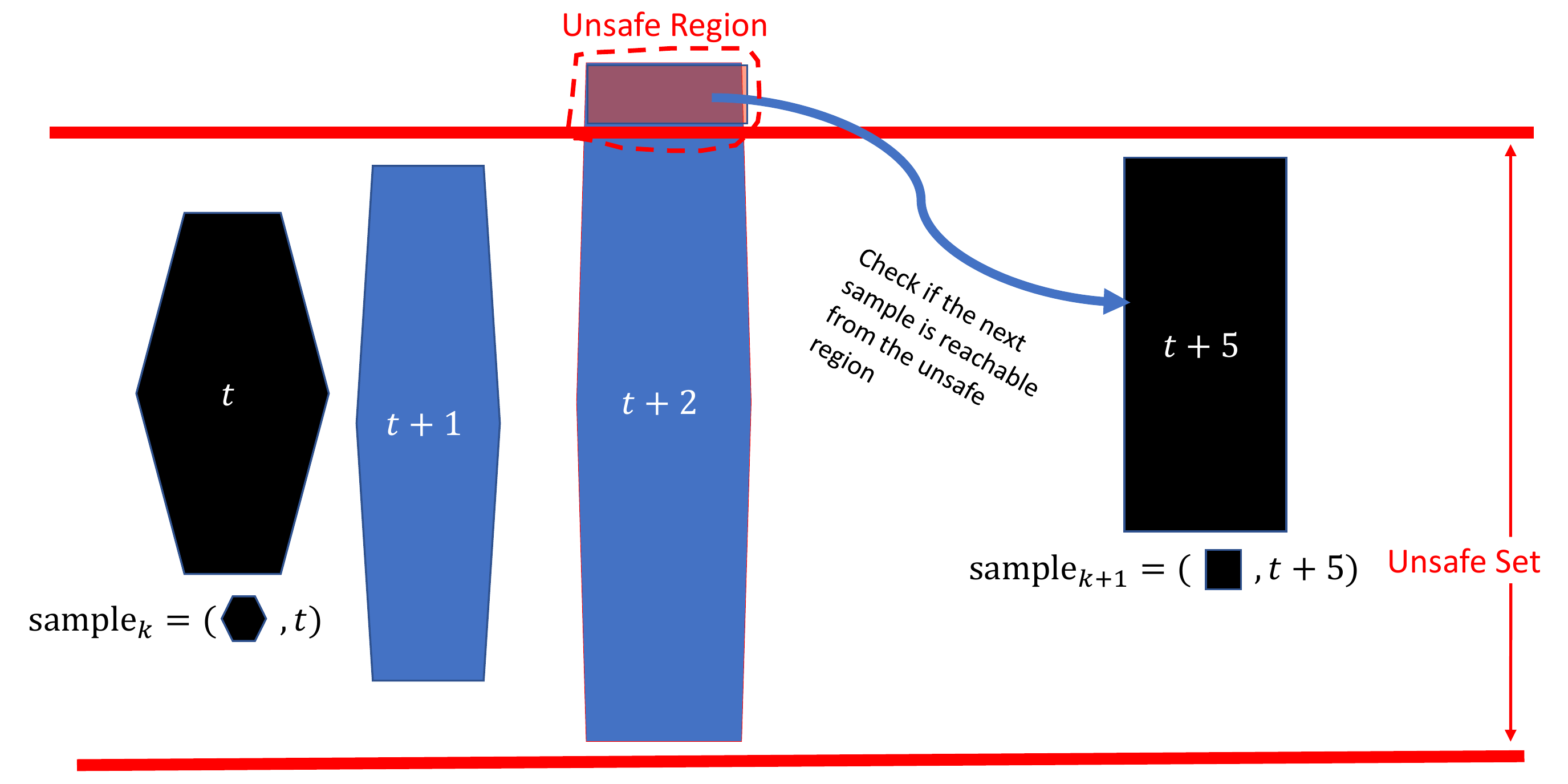}
        \vspace{-1.5em}
        \caption{\textbf{Offline}}
        \label{fig:offline_monitoring}
    \end{subfigure}
    \\
    \begin{subfigure}{.65\textwidth}
        \centering

        \includegraphics[width=\linewidth]{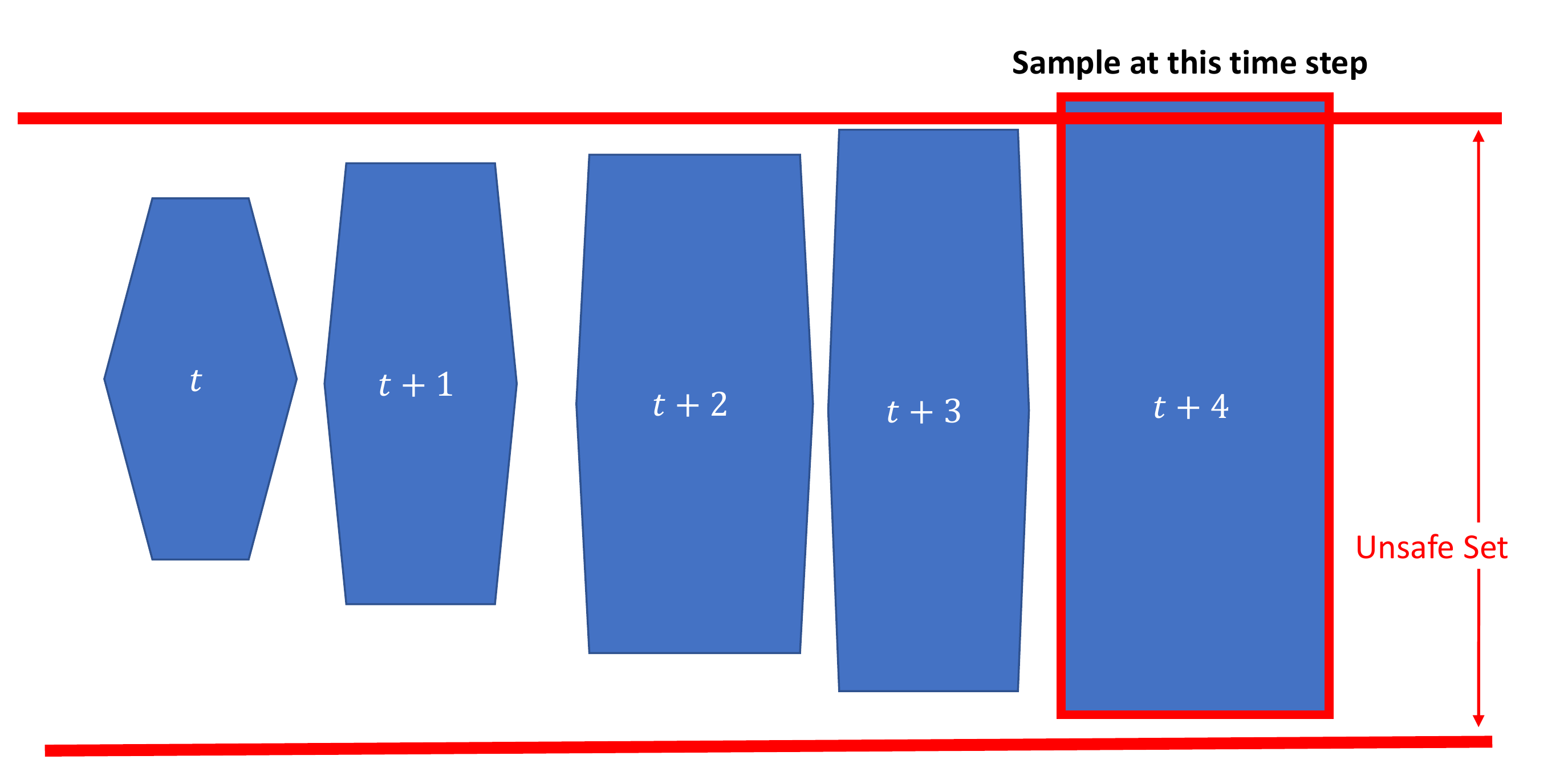}
        \vspace{-1.5em}
        \caption{\textbf{Online}}
        \label{fig:online_monitoring}
    \end{subfigure}
    \caption{(\ref{fig:offline_monitoring}): \textbf{Offline Monitoring}. Black: Two consecutive samples, $k$ and $k+1$, at time steps $t$ and $t+5$ respectively. Blue: The over-approximate reachable set computed from sample $k$ using \reach{.}.
    (\ref{fig:online_monitoring}): \textbf{Online Monitoring}. Blue: Over-approximate reachable set computed, at each step, using \reach{.}.}
    \label{fig:monitoring}
\end{figure*}

Our first contribution addresses offline monitoring: in this setting, we assume full knowledge of the (possibly scattered) uncertain log, usually after an execution is completely over.
In a first step, we assume that logs are known with full certainty regarding the time step; that is, the input is a \emph{fixed timestamp uncertain log}.
(The case of a fully uncertain log will be addressed in \cref{ss:offline-uncertain}.)

Before we propose our offline algorithm, we illustrate the approach in \cref{fig:offline_monitoring}.
Consider two consecutive samples $k$ and $k+1$, marked in black, at time steps $t$ and $t+5$ respectively. The reachable sets, in blue, represent the over-approximate behaviors possible by the system between time steps $t$ and $t+5$. Consider the case where at time step $t+2$ the over-approximate reachable set intersects with the unsafe region. Once our algorithm detects a possible unsafe behavior, it computes the intersection between the over-approximate reachable set (here, the reachable set at time-step $t+2$) and the unsafe set. Then it checks whether the reachable set, given in the next sample ($k+1$), is reachable from the unsafe region---if yes, it infers \emph{unsafe}; if not, it infers \emph{safe}.
Now, we formally propose our offline monitoring method in \cref{algo:offline} for a given fixed timestamp uncertain log~$\logg$.

\begin{algorithm}[tb]
\SetKwInOut{Input}{input}
\SetKwInOut{Output}{output}
\Input{An uncertain log $\logg$ of a system $x^+=\Lambda x$, and an unsafe set $\mathcal{U}$.}
\Output{Return \textit{safe} (resp.\ \textit{unsafe}) if the actual system behavior is safe (resp.\ potentially unsafe).}
\For{$k \in \{ 1 , \dots, |\logg|-1\}$ }
{
\nllabel{algo:offline:traverseSample}
  $\Big(\hat{\theta}_{t_k},\big[t^{lb}_k,t^{ub}_k\big]\Big) \leftarrow \logg_k$ \nllabel{algo:offline:sample_k} \tcp*{current sample, \HiLi[yellow!40]{with interval time step}}

  $\Big(\hat{\theta}_{t_{k+1}},\big[t^{lb}_{k+1},t^{ub}_{k+1}\big]\Big) \leftarrow \logg_{k+1}$ \nllabel{algo:offline:sample_k1} \tcp*{next sample, \HiLi[yellow!40]{with interval time step}}
  
  \For {\HiLi[yellow!40]{$t_k \in \big\{ t^{lb}_k , \dots, t^{ub}_{k}\big\}$} \nllabel{algo:offline:forallTk}}
  {
      \For {\HiLi[yellow!40]{$t_{k+1} \in \big\{ t^{lb}_{k+1} , \dots, t^{ub}_{k+1}\big\}$} \nllabel{algo:offline:forallTk2}}
      {
      $t_{\Delta}=t_{k+1}-t_k-1$ \tcp*{time gap between two possible time steps of the two samples}
      
      {\tcc{Compute reachable set for all possible time steps (from the two intervals of time steps) between two samples. Check if any of the sets intersect with the unsafe set}}

      \For {$p \in \{ 1 , \dots, t_{\Delta}-1\}$ \nllabel{algo:offline:forall}}
      {
        \If{$\hat{\theta}_{t_k+p} \cap \mathcal{U} \ne \emptyset$}{
        \nllabel{algo:offline:checkIntersection}
          {\tcc{Refinement process starts}}
          $\psi \leftarrow \hat{\theta}_{t_k+p} \cap \mathcal{U}$ \nllabel{algo:offline:compIntersection} \tcp*{compute the unsafe region of the system}
          $t_{d}=t_{k+1}-(t_k+p)$ \;
          $\vartheta \leftarrow \reach{\Lambda,\psi,t_d}$ \; \nllabel{algo:offline:refineRS}
          \tcc{Check if {the }next sample is reachable from {the }unsafe{ region}}
          \If{$\vartheta \cap \hat{\theta}_{t_{k+1}} \ne \emptyset$}{\nllabel{algo:offline:refineCheck}
        \Return{\textit{unsafe}}  \nllabel{algo:offline:returnUnsafe} \tcp*{{the }next sample is reachable from{ the} unsafe{ region}}
        }
        }
    
        $\hat{\theta}_{t_k+p+1} \leftarrow \reach{\Lambda,\hat{\theta}_{t_k+p},1}$ \;
      }

      }
  }

}
\Return{\textit{safe}} \;
\caption{Offline monitoring}
\label{algo:offline}
\end{algorithm}

\paragraph{Description}
As we first consider an input log given in the form of a fixed timestamp uncertain log (\cref{definition:fixed-log}), we consider a simpler version of \cref{algo:offline} without the highlighted lines:
at line~\ref{algo:offline:sample_k}, we have $t^{lb}_k = t^{ub}_k = t_k$, and at line~\ref{algo:offline:sample_k1} we have $t^{lb}_{k+1} = t^{ub}_{k+1} = t_{k+1}$;
in addition, the \textbf{for} loops at lines~\ref{algo:offline:forallTk} and~\ref{algo:offline:forallTk2} can be discarded.

Let us now describe this simpler version of the algorithm.
The \textbf{for} loop, starting in \cref{algo:offline:traverseSample}, traverses through each sample, and checks if the system can reach a possibly unsafe behavior between two consecutive samples (computed in \cref{algo:offline:sample_k,algo:offline:sample_k1}), using over-approximate reachable set computation.
If the over-approximate reachable set between two consecutive samples intersect with the unsafe set (\cref{algo:offline:checkIntersection}), we perform a refinement as follows (\cref{algo:offline:compIntersection}--\cref{algo:offline:returnUnsafe}): We compute the unsafe region (intersection between unsafe set and over-approximate reachable set) in \cref{algo:offline:compIntersection}, then check if we can reach the next sample from the unsafe region (\cref{algo:offline:refineRS}--\cref{algo:offline:returnUnsafe}). If the next sample is reachable from the unsafe behavior, we conclude the system is unsafe (\cref{algo:offline:refineCheck}--\cref{algo:offline:returnUnsafe}).

\paragraph{Soundness and incompleteness}
Our proposed offline monitoring approach is \emph{sound} at discrete time steps, but not \emph{complete}---there might be cases where our algorithm returns \emph{unsafe} even though the actual system is \emph{safe}.
The primary reason for its incompleteness is due to the fact that \reach{.} computes an over-approximate reachable set.
Formally:

\begin{restatable}[soundness at discrete time steps for a fixed timestamp uncertain log]{thm}{theoremofflinesafetyCT}
	\label{thm:offline:safetyCT}
If the actual system is unsafe at some discrete time step, then \cref{algo:offline} returns unsafe.
Equivalently, if \cref{algo:offline} returns safe, then the actual system is safe at every discrete {time }step.
\end{restatable}
\begin{proof}
	We consider a fixed timestamp uncertain log.
	Let the actual trajectory $\tau$, between two samples $k$ and $k+1$, become unsafe at time step $t_{un}$.
	Therefore, the over-approximate reachable set, computed by $\reach{\cdot}$ at time step $t_{un}$, will also intersect with the unsafe set (due to soundness of $\reach{\cdot})$. Note that the actual trajectory $\tau$, originating from the sample~$k$, intersects the unsafe region at time step $t_{un}$, and reaches the sample $k+1$.
	The refinement module (\cref{algo:offline}, \cref{algo:offline:compIntersection}--\cref{algo:offline:returnUnsafe}), using over-approximate reachable sets will therefore infer the same, concluding the system behavior to be unsafe.
\end{proof}
\subsection{Offline monitoring over uncertain logs}\label{ss:offline-uncertain}

We now extend our offline monitoring to logs with uncertainty not only in the state dimension, but also in the timestamp dimension (as in \cref{definition:log}).
The extended version of the algorithm is the full \cref{algo:offline}, including the highlighted parts.
We namely add a pair of \textbf{for} loops at lines~\ref{algo:offline:forallTk} and~\ref{algo:offline:forallTk2}, iterating over each (concrete) timestamp in the current uncertain sample ($[t^{lb}_k,t^{ub}_k]$) and over the next uncertain sample ($[t^{lb}_{k+1},t^{ub}_{k+1}]$).
That is, we handle uncertainty over the logging times by iterating over each possible concrete log time in the logged interval.
This is a crux to ensure soundness of our approach, and guarantee that a safe answer indeed guarantees safety of the actual system (at all discrete time steps).

Conversely, and as in \cref{ss:offline}, our algorithm is not necessarily \emph{complete} (our algorithm might return \emph{unsafe} even though the actual system is \emph{safe}) due to the over-approximation of the reachable set computation.

We prove formally the soundness of \cref{algo:offline} below:

\begin{restatable}[soundness at discrete time steps for an uncertain log]{thm}{theoremofflinesafety}
	\label{thm:offline:safety}
If the actual system is unsafe at some discrete uncertain time step, then \cref{algo:offline} returns unsafe.
Equivalently, if \cref{algo:offline} returns safe, then the actual system is safe at every discrete uncertain {time }step.
\end{restatable}
\begin{proof}
	Let the actual trajectory $\tau$, between two samples $k$ and $k+1$, with uncertain time steps $[t_k^{lb},t_k^{ub}]$ and $[t_{k+1}^{lb},t_{k+1}^{ub}]$, become unsafe at time step $t_{un}$.
	\cref{algo:offline} computes the reachable set of all possible time steps between $[t_k^{lb},t_k^{ub}]$ and $[t_{k+1}^{lb},t_{k+1}^{ub}]$.
	Therefore, the over-approximate reachable set, computed by $\reach{\cdot}$ at time step $t_{un}$, will also intersect with the unsafe set (due to soundness of $\reach{\cdot})$.
	Note that the actual trajectory $\tau$, originating from the sample $k$, intersects the unsafe region at time step $t_{un}$, and reaches the sample $k+1$.
	The refinement module (\cref{algo:offline}, \cref{algo:offline:compIntersection}--\cref{algo:offline:returnUnsafe}), using over-approximate reachable sets will therefore infer the same, concluding the system behavior to be unsafe.
\end{proof}
\subsection{Online monitoring over fixed timestamp uncertain logs}\label{ss:online}

We now move to \emph{online} monitoring.
In contrast to \cref{ss:offline-uncertain}, in our online setting, timestamps are necessarily exact, as we suppose we can trigger (instantaneously) a sample.
(Still, there could be cases where uncertainty in the timestamps could be useful in an online setting---this will be discussed in \cref{ss:future}.)
However, the logged states are still uncertain (as in \cref{definition:fixed-log}).
We propose our online monitoring method in \cref{algo:online}.

\begin{algorithm}
\SetKwInOut{Input}{input}
\SetKwInOut{Output}{output}
\Input{An uncertain system $x^+=\Lambda x$, an unsafe set $\mathcal{U}$, time bound $H$.}
\Output{Return \textit{safe} \textit{iff} the actual system behavior is safe.}
$\hat{\theta}_0 \leftarrow \text{Sampling at time step 0}$ \nllabel{algo:online:initSample} \tcp*{initial behavior of the system.}
\tcc{Check whether the initial behavior is safe}
\lIf{$\hat{\theta}_{0} \cap \mathcal{U} \ne \emptyset$}{\nllabel{algo:online:initIntersection}
  \Return{\textit{unsafe} \nllabel{algo:online:initReturnUnsafe}}}

\For{$t \in \{ 1, 2, \dots, H-1\}$ \label{algo:online:for}}
{
  $\hat{\theta}_{t+1} \leftarrow \reach{\Lambda,\hat{\theta}_{t},1}$ \tcp*{over-approximate reachable set{ at next step}}

  \tcc{Check whether the over-approximate reachable set is unsafe}
  \If{$\hat{\theta}_{t+1} \cap \mathcal{U} \ne \emptyset$}{ \nllabel{algo:online:checkIntersection}
  $\logg_{t+1} \leftarrow \text{Sample at time step }t+1$ \nllabel{algo:online:sampleT1} \;
  \tcc{Check whether the actual reachable set is unsafe}
  \If{$\logg_{t+1} \cap \mathcal{U} \ne \emptyset$}{
  \Return{\textit{unsafe}} \nllabel{algo:online:returnUnsafe}\;
  }
  $\hat{\theta}_{t+1} = \logg_{t+1}$ \nllabel{algo:online:reset} \tcp*{reset to actual behavior}
  }
}
\Return{\textit{safe}};
\caption{Online monitoring}
\label{algo:online}
\end{algorithm}

\paragraph{Description}
The online monitoring algorithm begins by sampling the system at the initial time step, say 0, in \cref{algo:online:initSample}. As a sanity check, we confirm if the initial behavior of the system is safe in \cref{algo:online:initIntersection}.
The \textbf{for} loop starting in \cref{algo:online:initIntersection}---where each iteration corresponds to the set of actions for a time step~$t$---performs the following: At a given time step $t$, we compute the over-approximate reachable set at the next time step $t+1$ (\cref{algo:online:sampleT1}). If the computed over-approximate reachable set intersects with the unsafe set, we sample the system at time step $t+1$ to check if the actual behavior is also unsafe (\cref{algo:online:checkIntersection}--\cref{algo:online:reset}). If safe, we reset the behavior (\cref{algo:online:reset}); if unsafe, we return \textit{unsafe} (\cref{algo:online:returnUnsafe}). Intuitively, this method samples the actual system only when the over-approximate reachable set, computed by \reach{.}, intersects the unsafe set.
This process is illustrated in \cref{fig:online_monitoring}.

\paragraph{Soundness and completeness}
Our online monitoring algorithm is correct (safe and complete) at discrete time steps, \emph{provided} the samples are accurate---it returns safe if and only if the actual behavior of the system is safe at all discrete time steps, when accurate samples are obtained.
Intuitively, we get the completeness from the fact that it returns unsafe if and only if the (accurate) sample is unsafe. Formally:

\begin{restatable}[correctness at discrete time steps]{thm}{theoremonlinesafety}
	\label{thm:online:safety}
\cref{algo:online} returns safe iff the actual behavior at all discrete time steps is safe.
\end{restatable}
\begin{proof}
	The soundness proof---if the actual behavior is unsafe, \cref{algo:online} infers unsafe---is straightforward.
	Hence, we now argue the completeness---if the actual behavior is safe, \cref{algo:online} infers \emph{safe}.
	Note that, \cref{algo:online} infers the system behavior as \emph{unsafe} only when a sampled log (actual behavior) becomes unsafe: therefore, if the samples are free from uncertainties (\ie{} exact), \cref{algo:online} is complete.
\end{proof}

\begin{rem}\label{remark:precision}
	While our aim is to consider continuous systems, note that, for \emph{discrete-time systems}, our approach is entirely correct (sound and complete), without the need for a restriction to ``discrete time steps'', since we can find a granularity small enough for the discrete-time evolution.
	This is notably the case for systems where the behavior does not change faster than a given frequency (\eg{} the processor clock).
	In the case of controllers, the granularity can be chosen by selecting the sampling period (the period at which a control input is applied).
\end{rem}

\begin{rem}\label{rem:scalingRS}
Given our reliance on enumerating time steps in both offline and online monitoring approaches, an increase in the granularity of sampling periods will necessitate the computation of a larger number of reachable sets.
Consequently, this may slow down the analysis process. Nonetheless, the reachable set computation method employed in this work exhibits excellent scalability when dealing with small time steps, as shown in~\cite{ghosh1}.
In other words, if the time interval between two samples in a log is not significantly large (\eg{} an order of 500~steps), this technique can easily compute reachable sets.
Moreover, if the time gaps do exceed 500~steps, one can enhance the scalability of the reachable set computation by utilizing the interval-based or zonotope-based reduction methods proposed in~\cite[Section~5.2]{ghosh1}.
\end{rem}

We will study the scalability of our approach in the next section.

\section{Case studies}\label{sec:case_studies}

We demonstrate the applicability and usability of our approach on three benchmarks: a medical device (\cref{subsec:anaesthesia}), an adaptive cruise control (\cref{subsection:ACC}), and an aircraft orbiting system (\cref{subsection:aircraft}).

\subsection{Implementation and environment}

We implemented our offline (both using a fixed timestamp uncertain log and a fully uncertain log) and online monitoring algorithms in a Python-based prototype tool, named \MoULDyS{}~\cite{GHOSH2023102976}.
Tool source and binaries, models and raw results are publicly available on GitHub\footnote{\url{https://github.com/bineet-coderep/MoULDyS/tree/uncertain_timestamp}}.
Further, the results in this paper can be easily recreated using the scripts provided in the Github repository\footnote{\url{https://github.com/bineet-coderep/MoULDyS/tree/uncertain_timestamp/src/recreate_results_from_paper}} and the reproducible artifact\footnote{%
	\url{https://zenodo.org/doi/10.5281/zenodo.7888501}
}.

\paragraph{Experimental environment}
All our experiments were performed on a Lenovo ThinkPad Mobile Workstation with i7-8750H CPU with 2.20 GHz and 32\,GiB memory on Ubuntu 20.04 LTS {operating system }(64 bit).
Our tool uses \texttt{numpy}~\cite{numpy}, \texttt{scipy}~\cite{scipy}, \texttt{mpmath}~\cite{mpmath} for matrix multiplications, \cite{ghosh1} to compute \reach{.}, and the \texttt{Gurobi}~\cite{gurobi} engine for visualization of the reachable sets.

\paragraph{Implementation details vis-à-vis \cref{algo:offline,algo:online}}
The intersection checking between two sets in \cref{algo:offline,algo:online} has been implemented as an optimization formulation in \texttt{Gurobi}.
That is, given two sets, our implementation of intersection check returns true iff the two sets intersect.
In other words, our intersection check is \emph{exact}.
In contrast, \emph{computing} the result of the intersection between two sets adds an over-approximation in our implementation. Given two sets, we compute a box hull of the two sets and then compute intersection of the two box hulls. Therefore, the only over-approximate operation we perform in \cref{algo:offline,algo:online}---apart from $\reach{\cdot}$---is \cref{algo:offline:compIntersection} in \cref{algo:offline}.

\paragraph{Generating scattered uncertain logs for offline monitoring}
At each time step, the logging system may take a snapshot of the system evolution at that time step; the logging occurs with a probability~$p$ (given).
However, it is impossible to determine the precise timestamp of the log if it is being transmitted across a shared network or if the clock tracking the time has errors. Instead, the timestamp then changes into an interval that contains all potential timestamps---this can be referred to as an uncertain timestamp.
Given a possible timing delay of $t_\delta$ (as per the network's quality), the size of the interval representing the timestamp associated with the log can be anywhere between 0 to $t_\delta$ (not necessarily exactly equal to $t_\delta$).
In other words, at each such uncertain timestamp, it records the evolution of the system with probability~$p$.
Clearly, due to the probabilistic logging, this logger is not guaranteed to generate periodic samples.
In each of our three case studies, it is important to highlight that the state variables under monitoring encompass the following: 
\begin{ienumerate}
    \item concentration levels within the anesthesia system,  
    \item distance, acceleration, and velocity within an ACC model,
    \item positional data of an aircraft.
\end{ienumerate}
It is evident that in practical scenarios, it becomes nearly impossible to measure the precise values of these variables due to the inherent susceptibility of the sensors used to errors. Nonetheless, considering a specific error margin for a sensor, it becomes feasible to calculate the uncertain value of the variables by adjusting the error tolerance of the sensor.
Consequently, we also do not assume that the samples logged by the logging system, at each timestamp, are accurate---the logging system, due to sensor uncertainties, logs an over-approximate sample of the system at that time step.
In our experiments, each uncertain log was generated statically from our bounding model (the uncertain linear dynamical system) by simulating its evolution from an uncertain initial set (\ie{} not reduced to a point).
In the end, we get an uncertain log (as in \cref{definition:log}).

For the first two out of three case studies, we chose two values for the logging probability~$p$ of 20\,\% and~40\,\% respectively.
They were selected empirically, as our experiments showed that these two values led to quite different behaviors for our offline algorithm: 40\,\% can be considered as a frequent sampling, while 20\,\% is more sporadic.
We used the same probabilities throughout these two case studies to allow for fair comparison, and for general observations on the effect of sampling probability and uncertainty across experiments.\label{newtext:choosing-p}

\paragraph{Logging system for online monitoring}
When the logging system is triggered, at a time step, to generate a sample, the logging system records the evolution of the system and sends it to the online monitoring algorithm.
Similar to the offline logging system, we do not assume that the samples logged by the logging system are perfectly accurate{---the logging system, due to sensor uncertainties, logs an over-approximate sample of the system at that time step.
That is, we use the same method---as the offline logging system---to generate logs (statically), but unlike the offline algorithm, the online algorithm uses the samples only when required}.
For all our case studies, all the generated logs are \emph{safe}.

\paragraph{Research questions}
We consider the following research questions{ in our case studies}:

\begin{oneenumerate}
    \item Effect of logging probability (number of log samples) on the rate of false alarms raised by the offline monitoring---inferring a behavior as ``potentially unsafe'' when the actual behavior is ``safe''.

    \item For offline monitoring, does the size of the samples (in other words, volume of the set obtained as sample), gathered at each step, have an impact on the rate of false alarms?
    Put it differently, what is the effect, \emph{vis-à-vis} false alarms, of the amount of the uncertainty in the log?

    \item For online monitoring, how frequent is the logging system triggered to generate a sample?

    \item For the same execution, how do the outcome (in terms of verdict on safety by the monitoring algorithms) and the efficiency (in terms of number of samples needed) of the offline and online monitoring algorithms compare?
    
    \item The effect of timing uncertainty on the rate of false alarms raised by the offline monitoring?
\end{oneenumerate}

Using the first two case studies, automated anesthetic delivery and adaptive cruise control, Questions~(1)-(4) will be addressed.
The airplane orbiting benchmark will be used to provide the answer to Question~(5).
This design decision was made because, unlike the other two benchmarks, only the airplane orbiting benchmark transmits logs through a shared network; in the other two examples, logs are collected locally over a reliable channel.

\subsection{First benchmark: Anesthesia}\label{subsec:anaesthesia}
\subsubsection{System description}

\begin{figure*}
    \begin{subfigure}{.48\textwidth}
        \centering
        \includegraphics[width=\linewidth]{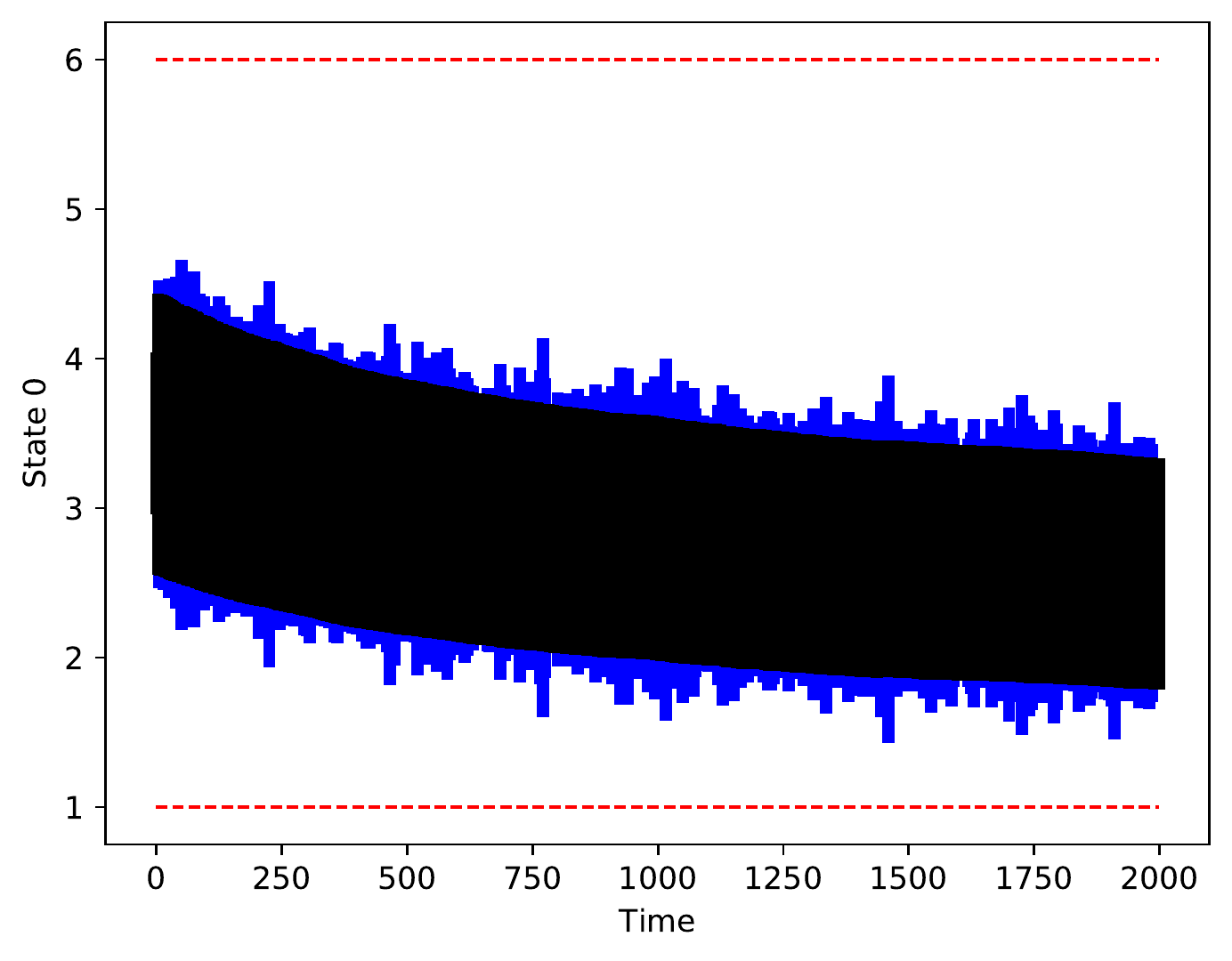}
        \caption{Frequent samples, low uncertainty}
        \label{fig:anesthesia_fig3}
    \end{subfigure}
    \hfill
    \begin{subfigure}{.48\textwidth}
        \centering
        \includegraphics[width=\linewidth]{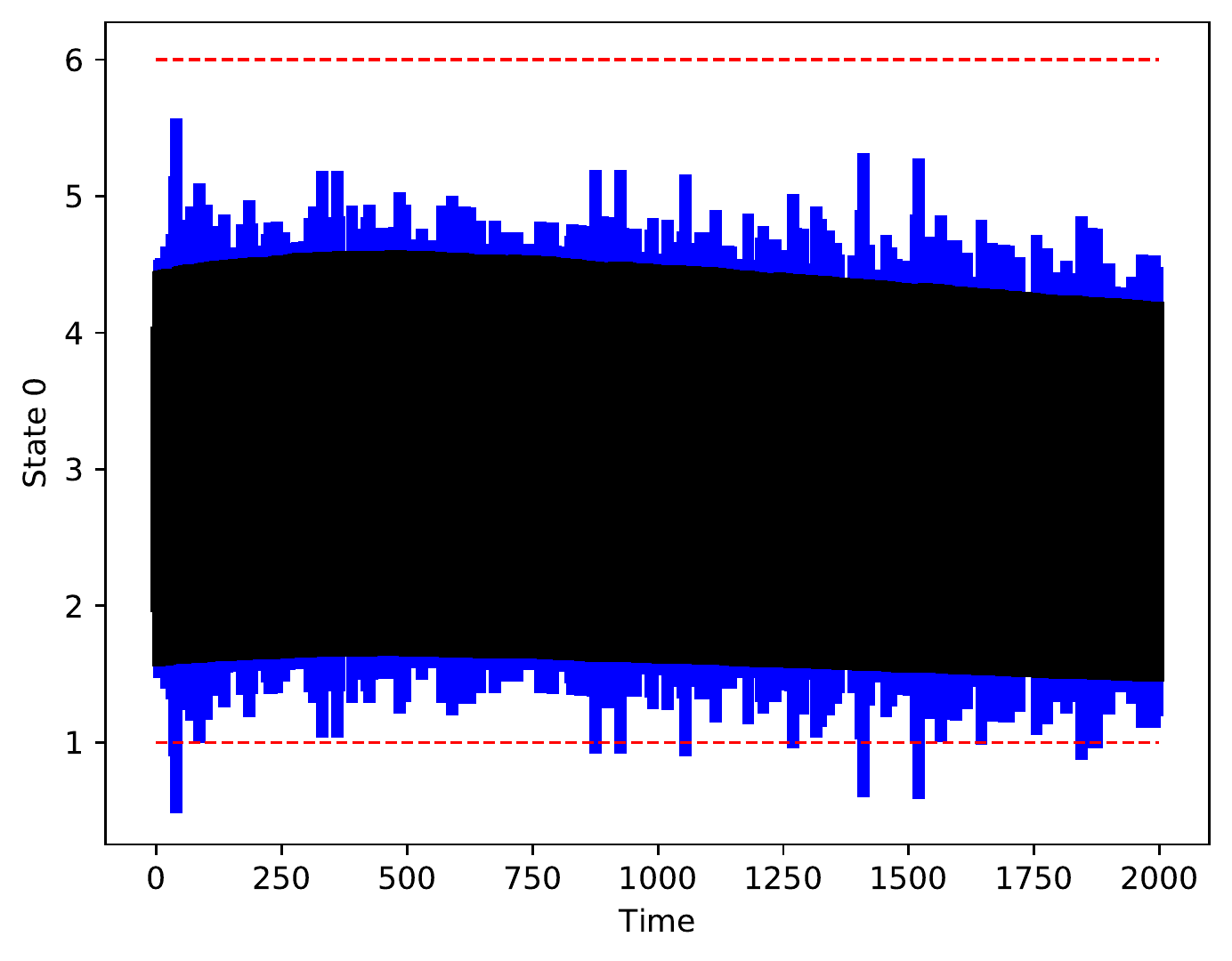}
        \caption{Frequent samples, high uncertainty}
        \label{fig:anesthesia_fig4}
    \end{subfigure}
    
    \vfill
    
    \begin{subfigure}{.48\textwidth}
        \centering
        \includegraphics[width=\linewidth]{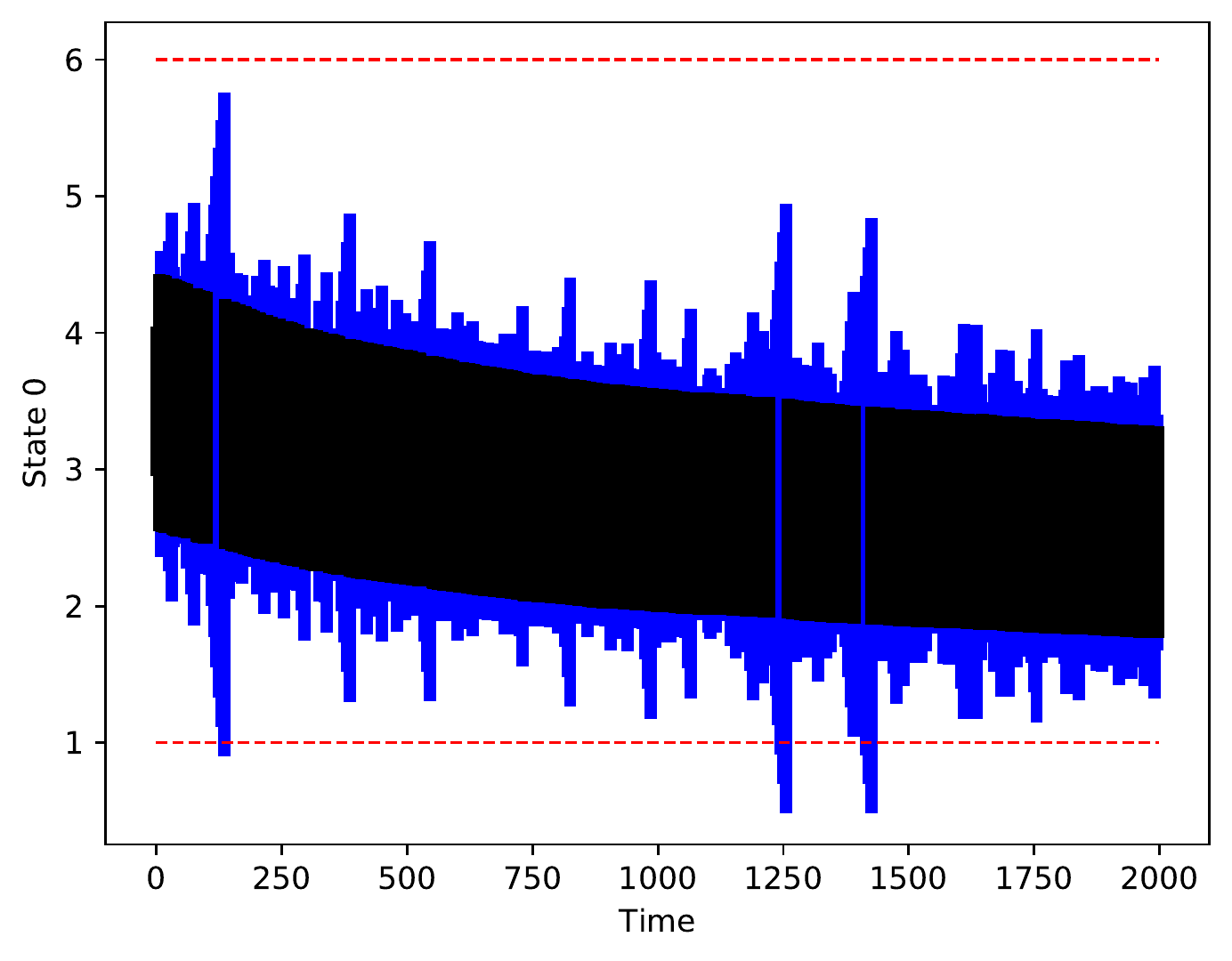}
        \caption{Sporadic samples, low uncertainty}
        \label{fig:anesthesia_fig1}
    \end{subfigure}
    \hfill
    \begin{subfigure}{.48\textwidth}
        \centering
        \includegraphics[width=\linewidth]{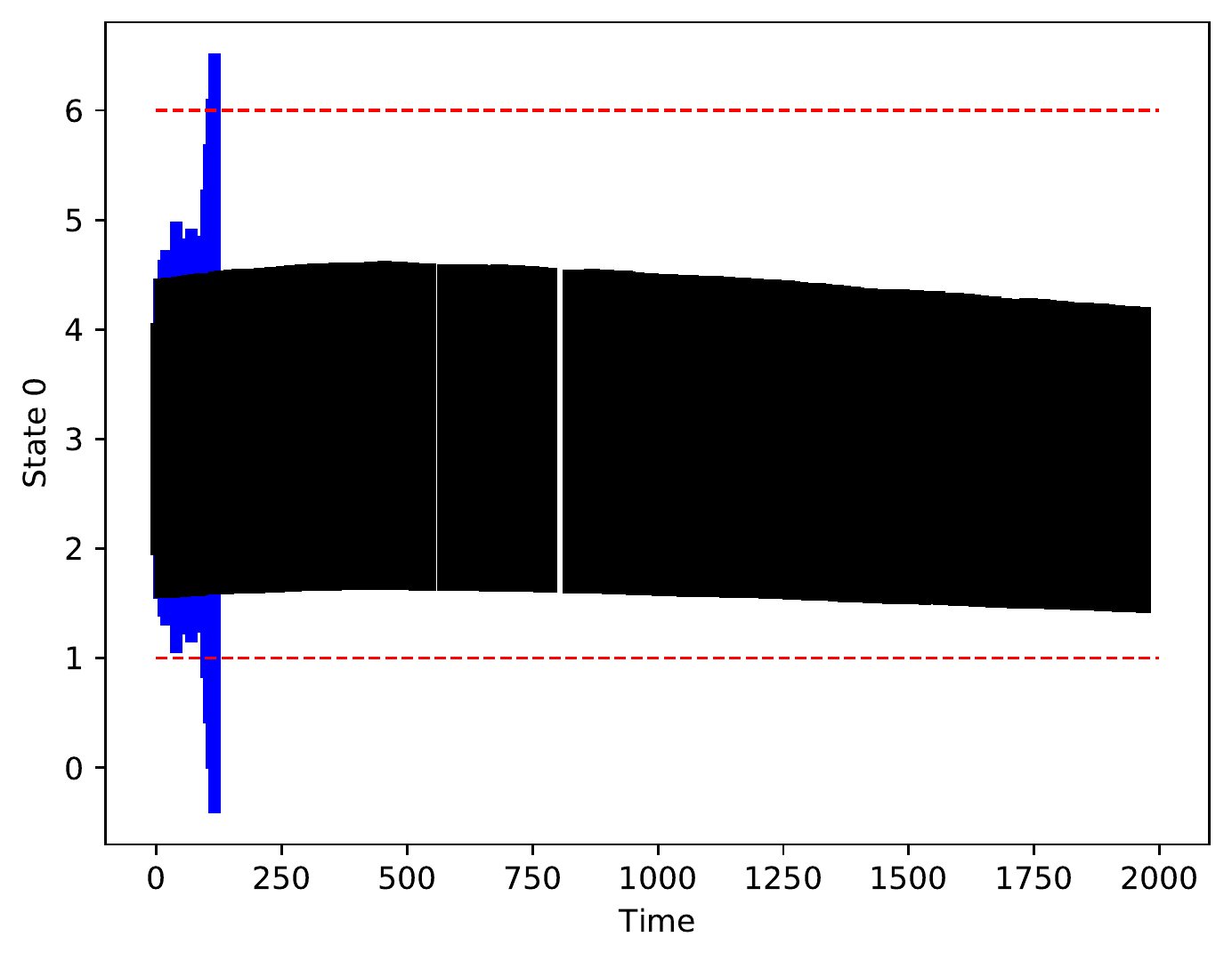}
        \caption{Sporadic samples, high uncertainty}
        \label{fig:anesthesia_fig2}
    \end{subfigure}
    \caption{\textbf{\textit{Offline Monitoring (Anesthesia).}} We plot the change in concentration level of $c_p$ with time. The volume of the samples increases from left to right, and the probability of logging increases from bottom to top. The blue regions are the reachable sets showing the over-approximate reachable sets as computed by the offline monitoring, the black regions are the samples from the log given to the offline monitoring algorithm, and the red dotted line represents safe distance level. Note that although \cref{fig:anesthesia_fig1,fig:anesthesia_fig4} reachable sets' seem to intersect with the red line (unsafe set), the refinement module infers them to be \emph{unreachable}, therefore concluding the system behavior as \emph{safe}---unlike \cref{fig:anesthesia_fig2}. } 
    \label{fig:offline}
\end{figure*}

\begin{figure*}
    \begin{subfigure}{.49\textwidth}
        \centering
        \includegraphics[width=\linewidth]{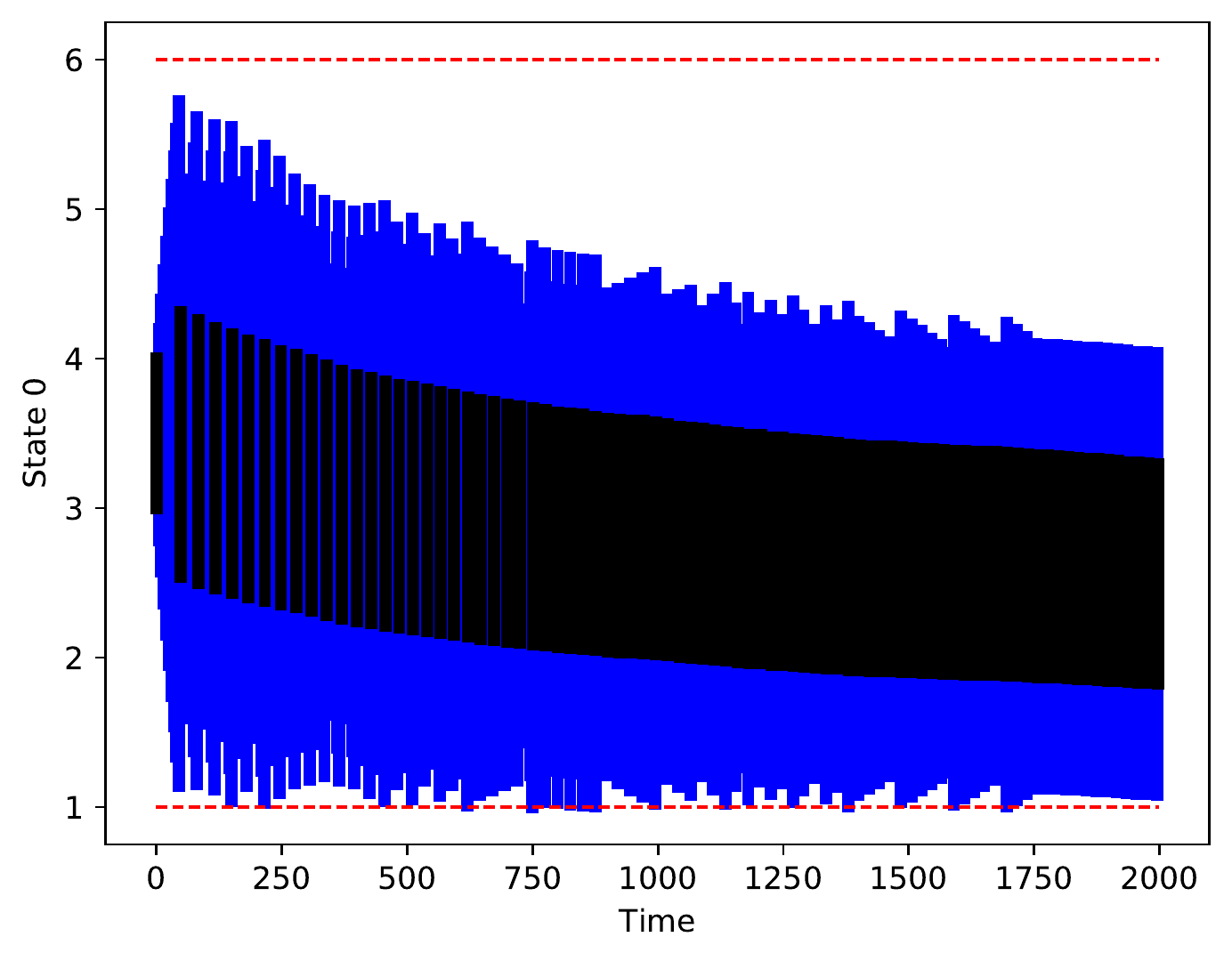}
        \caption{\textbf{Online monitoring}}
        \label{fig:anesthesia_online}
    \end{subfigure}
    \hfill
    \begin{subfigure}{.49\textwidth}
        \centering
        \includegraphics[width=\linewidth]{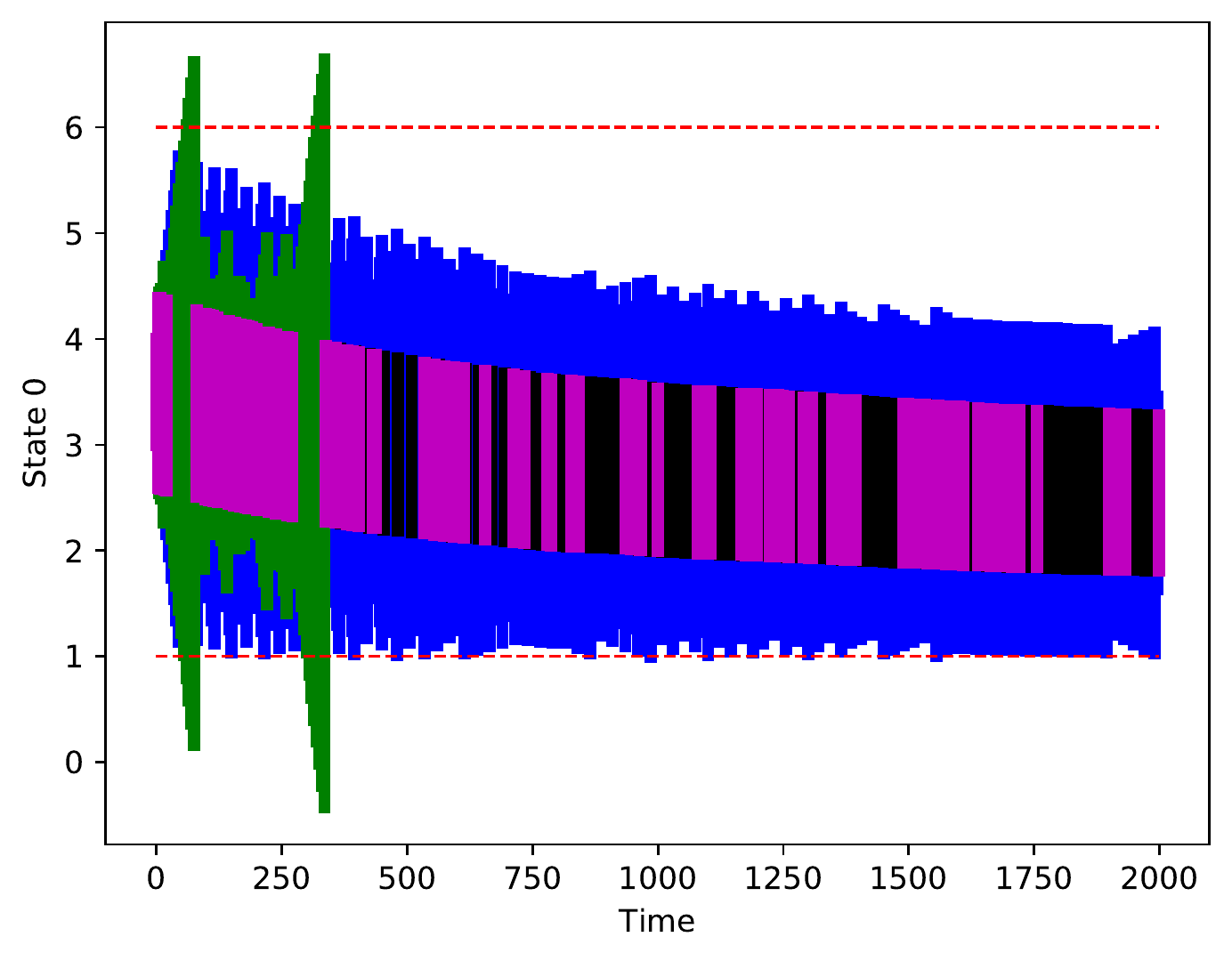}
        \caption{\textbf{Comparing online and offline monitoring}}
        \label{fig:anesthesia_comp}
    \end{subfigure}
    \caption{\textbf{\textit{Online Monitoring (Anesthesia).}} We plot the change in concentration level of~$c_p$ with time. The blue regions are the reachable sets showing the over-approximate reachable sets as computed by the online monitoring, the black regions are the samples generated when the logging system was triggered by the online monitoring algorithm, and the red dotted line represents safe concentration levels. \textit{Online monitoring (\cref{fig:anesthesia_online}):} We apply our online monitoring to the anesthesia model. \textit{Comparison (\cref{fig:anesthesia_comp}):} We compare our online and offline algorithms. The green regions are the reachable sets showing the over-approximate reachable sets between two consecutive samples from the offline logs, the magenta regions are the offline logs, given as an input to the offline monitoring algorithm, generated by the logging system, and the red dotted line represents safe concentration levels. The blue regions are the reachable sets showing the over-approximate reachable sets as computed by the online monitoring, the black regions are the samples generated when the logging system was triggered by the online monitoring algorithm, and the red dotted line represents safe concentration levels.}
    \label{fig:online}
\end{figure*}

We first demonstrate our approach on an automated anesthesia delivery model~\cite{ARCH15:Benchmark_Problem_PK_PD_Model}. The anesthetic drug considered in this model is propofol. Such safety critical systems are extremely important to be verified formally before they are deployed,
as under or overdose of the anesthetic drug can be fatal to the patient.

\paragraph{Model} The model as in~\cite{ARCH15:Benchmark_Problem_PK_PD_Model} has two components:
\begin{oneenumerate}%
	\item Pharmacokinetics (PK): models the change in concentration of the drug as the body metabolizes it.
	\item Pharmacodynamics (PD): models the effect of drug on the body.
\end{oneenumerate}%
The PK component is further divided into three compartments:
\begin{ienumerate}%
	\item first peripheral compartment~$c_1$,
	\item second peripheral compartment~$c_2$,
	\item plasma compartment~$c_p$.
\end{ienumerate}%
The PD component has one compartment, called~$c_e$. The set of state variables of this system is
$
[\begin{smallmatrix}
c_p & c_1 & c_2 & c_e
\end{smallmatrix}]^\top
$. The input to the system is the infusion rate of the drug (propofol)~$u$.
The complete state-space model of this system in given in~\cite[Equation~5]{ARCH15:Benchmark_Problem_PK_PD_Model}.

\paragraph{Model parameters}\label{paragraph:modelparameters}
The evolution of state variables $c_p$, $c_1$, $c_2$ is dependent on a number of parameters, such as: the weight of the patient ($\mathit{weight}$), and a number of ``first order rate constants'' between the compartments (called $k_{10}$, $k_{12}$, $k_{13}$, $k_{21}$ and~$k_{31}$ in~\cite{ARCH15:Benchmark_Problem_PK_PD_Model}, where their value is given).
The evolution of the fourth state variable~$c_e$ is dependent on the rate constant between plasma and effect site (called parameter~$k_d$ in~\cite{ARCH15:Benchmark_Problem_PK_PD_Model}).

\paragraph{Safety}
The system is considered safe (as suggested in \cite{ARCH15:Benchmark_Problem_PK_PD_Model}) if the following concentration levels are maintained at all time steps: $c_p \in [1,6]$, $c_1 \in [1,10]$, $c_2 \in [1,10]$, $c_e \in [1,8]$.
Note that this safety property is a hypercube, \ie{} a simple form of a zonotope.

In this case study, we focus our attention on the effect of perturbation, in the weight of the patient ($\mathit{weight}$), on the concentration level of plasma compartment~$c_p$.
Only the weight of the patient is subject to perturbation.
We assume that the weight of the patient has an additive perturbation of $\pm 0.8$\,kg in this case study---at each time step, the weight of the patient is $\mathit{weight} + \delta_w$, $\delta_w \in [0,0.8]$. With perturbation in the weight, we want to infer safety of this system using monitoring.

Clearly, monitoring of this system \textit{vis-à-vis} safety is crucial.
It is not practical for a busy human doctor or a practitioner to monitor each patient continuously at all time steps---monitoring, either offline or online, provides them an efficient way to save their time and treat their patients without compromising their safety.

The logs for this case study are uncertain ``only'' in the ``valuation'' dimension: that is, the logged valuation are known only with a finite precision, but the timestamps are exact; in other words, the input logs are fixed timestamp uncertain logs.

\subsubsection{Experiments}\label{anesthesia:experiments}

We now answer questions (1)-(4), using \cref{fig:offline,fig:online}.
In \cref{fig:offline}:
\begin{ienumerate}%
	\item the plots in the bottom row have logging probability of 20\,\%, and the plots in top row have a logging probability of~40\,\%;
	\item the plots in left column and the right column have been simulated with an initial set of $[\begin{smallmatrix}
[3,4] & [3,4] & [4,5] & [3,4]
\end{smallmatrix}]^\top
$, $u \in [2,5]$ and
$
[\begin{smallmatrix}
[2,4] & [3,6] & [3,6] & [2,4]
\end{smallmatrix}]^\top
$, $u \in [2,10]$
respectively.
\end{ienumerate}%
That is, the volume of the samples increases from left to right. In \cref{fig:online}, we simulated the trajectory with an initial set $[\begin{smallmatrix}
[3,4] & [3,4] & [4,5] & [3,4]
\end{smallmatrix}]^\top
$, $u \in [2,5]$.

\paragraph{Initial state}\label{anesthesia:initial}
The initial set is chosen such that the system starts from a safe specification.
For each generated log, we %
start from the initial set,
and add the uncertainty depending on our experimental context.

\paragraph{Answer to Question~1}
    We answer this question by comparing two sets figures in the left column (\cref{fig:anesthesia_fig1,fig:anesthesia_fig3}) and the right column (\cref{fig:anesthesia_fig2,fig:anesthesia_fig4}) of \cref{fig:offline}. \textit{For the left column, \ie{} with smaller sample size}: \cref{fig:anesthesia_fig1} took 51.40\,s and concluded the system to be safe. The analysis in this plot invoked the refinement module of the offline algorithm. But increasing the probability of logging, \ie{} more number of samples, as in \cref{fig:anesthesia_fig3}, resulted in not invoking the refinement module at all, thus taking 32.92\,s. \textit{For the right column, \ie{} with larger sample size}: this analysis, as shown in \cref{fig:anesthesia_fig2}, took 1.73\,s to complete, and concluded the system behavior to be unsafe. The behavior of the system, shown in \cref{fig:anesthesia_fig4} with 40\,\% probability of logging, results in inferring the behavior of the system as safe, by invoking the refinement module several times. Overall, this analysis, as shown in \cref{fig:anesthesia_fig4}, took 35.93\,s to complete, and concluded the system behavior to be safe.

\paragraph{Answer to Question~2}
    We answer this question by comparing two sets figures in the top row (\cref{fig:anesthesia_fig3,fig:anesthesia_fig4}) and the bottom row (\cref{fig:anesthesia_fig1,fig:anesthesia_fig2}) of \cref{fig:offline}. \textit{For the bottom row, \ie{} with smaller logging probability}: Increasing the volume of the samples results in inferring the behavior from safe (\cref{fig:anesthesia_fig1}) to unsafe (\cref{fig:anesthesia_fig2}), as per the offline monitoring algorithm. \textit{For the top row, \ie{} with higher logging probability}: Increasing the volume of the samples results in not invoking the refinement module (\cref{fig:anesthesia_fig3}) to invoking the refinement module several times (\cref{fig:anesthesia_fig4}), as per the offline monitoring algorithm.

\paragraph{Answer to Question~3}
    The result is given in \cref{fig:anesthesia_online}. Using our online algorithm, we were able to prove safety of the system in 109.04\,s.
    The online algorithm triggered the logging system to generate samples for 83~time steps---this is less than 5\,\% of total time steps.
    We observe, as shown in \cref{fig:anesthesia_online}, that the logging system is triggered more when the trajectory is closer to the unsafe region.

\paragraph{Answer to Question~4}
	We compare our offline and online algorithms, for $2\,000$ time steps, on the same trajectory. The result is given in \cref{fig:anesthesia_comp}. Note that, using our online algorithm, we were able to prove safety of the system in 107.99\,s. The online algorithm triggered the logging system to generate samples only 84 times. In contrast, the offline algorithm, with a log size of~115 (5\,\% logging probability) stopped at the 35th sample, (wrongly) inferring the system as unsafe, taking 71.37\,s.
\subsection{Second benchmark: Adaptive Cruise Control}\label{subsection:ACC}
We now apply our {offline and online monitoring }algorithms to an adaptive cruise control (ACC) model~\cite{acc}.

\subsubsection{System description}

An {adaptive cruise control} behaves like an ordinary cruise control when there is no car in the sight of its sensor, and when there is a car in its sight, it maintains a safe distance.

\paragraph{Model}
The model as in \cite{acc} has the following state variables:
\begin{ienumerate}%
	\item velocity of the vehicle~$v$,
	\item distance between the two vehicles~$h$, and
	\item velocity of the lead vehicle $v_L$.
\end{ienumerate}%
The state space of the system is given in~\cite[Equation~3]{acc}.
The set of state variables of this system is
$
[\begin{smallmatrix}
v & h & v_L
\end{smallmatrix}]^\top
$.

\paragraph{Model parameters}\label{ACC:parameters}
The model is dependent on two parameters:
\begin{ienumerate}%
	\item acceleration of the lead vehicle $a_L$, and
	\item breaking force and torque applied to the wheels as a lumped net force~$F$.
\end{ienumerate}%
Note that the model is dependent on acceleration of the vehicle~$a_L$, which is very hard to accurately measure due to sensor uncertainties. Similarly the torque~$F$ applied to the wheels is also dependent of the coefficient of friction of the ground.
To reflect such uncertainties, we consider $a_L \in [-0.9,0.6]$ and $F \in [-0.6,2.46]$.
These parameters were chosen as per~\cite[Tab.~1, Eq.~(6)]{acc}.

\paragraph{Safety}
We selected the following safety constraint:
The system is considered safe if the distance between vehicles $h > 0.5$.
(The unit is, as in~\cite{acc}, meters.)

Consider an event of a car crash, where the log stored by the car before the crash, is the only data available to analyze the crash; such an analysis might benefit police, insurance companies, vehicle manufacturers, etc.
Using our offline algorithm one can figure out if the car might have shown unsafe behavior or not.
Similarly, consider a vehicle on a highway with a lead vehicle in its sight.
The ACC in such a case needs to continuously read sensor values to track several parameters, such as acceleration of the lead vehicle, braking force, etc.---this results in wastage of energy.
Using our online monitoring algorithm, the car reads sensor values only when there is a potential unsafe behavior.
This intermittent behavior will result in saving energy without compromising safety of the system.

With the aforementioned reasons for applying our offline and online monitoring, we apply our algorithms on the ACC model and answer questions (1)-(4).
We think that the answers to these questions will help the car designers to design efficient ACC models without compromising safety.
Again, the logs for this second case study are uncertain ``only'' in the ``valuation'' dimension: that is, they are fixed timestamps uncertain logs.

\begin{figure*}
    \begin{subfigure}{.48\textwidth}
        \centering
        \includegraphics[width=\linewidth]{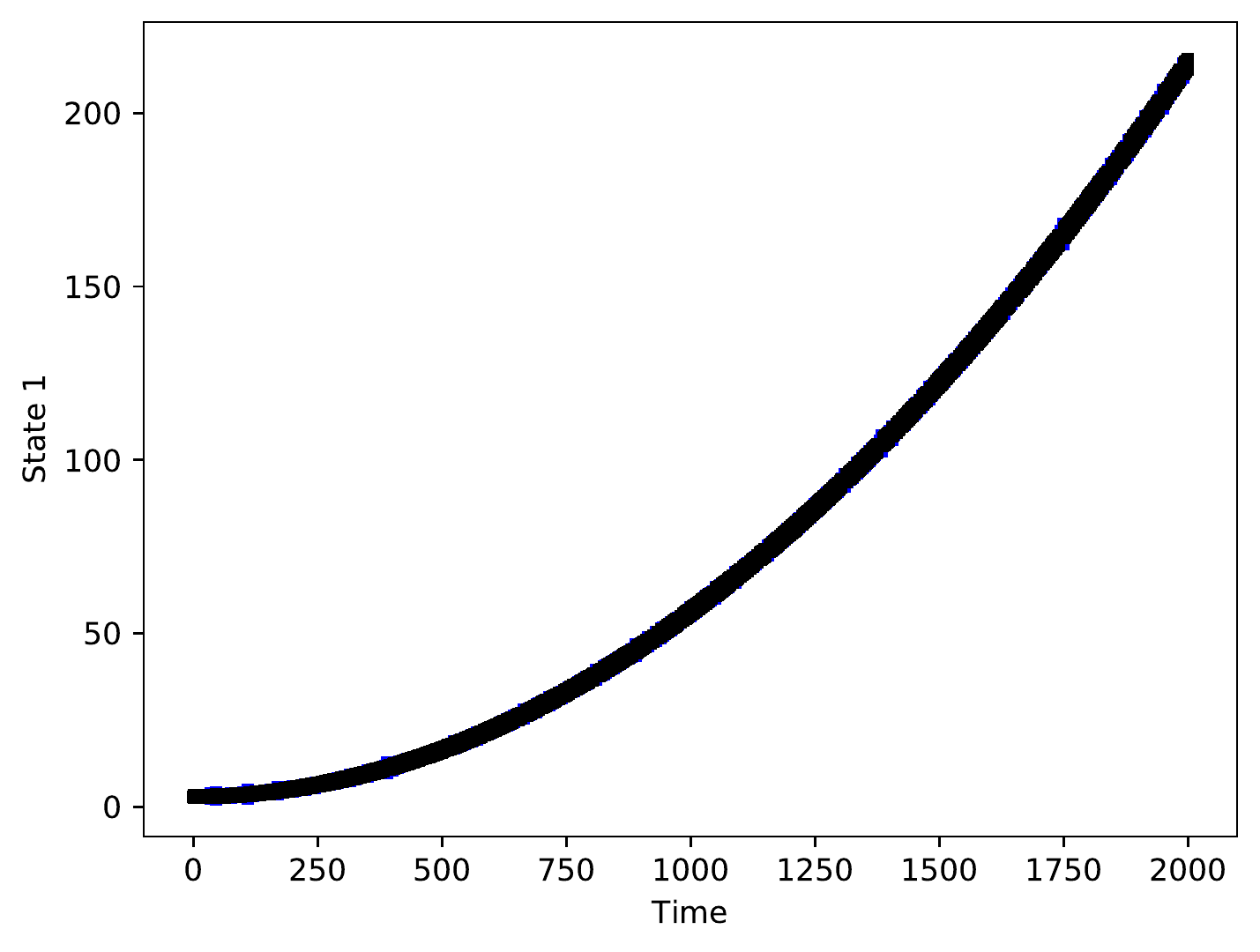}
        \caption{Frequent samples, low uncertainty}
        \label{fig:acc_fig3}
    \end{subfigure}
    \hfill
    \begin{subfigure}{.48\textwidth}
        \centering
        \includegraphics[width=\linewidth]{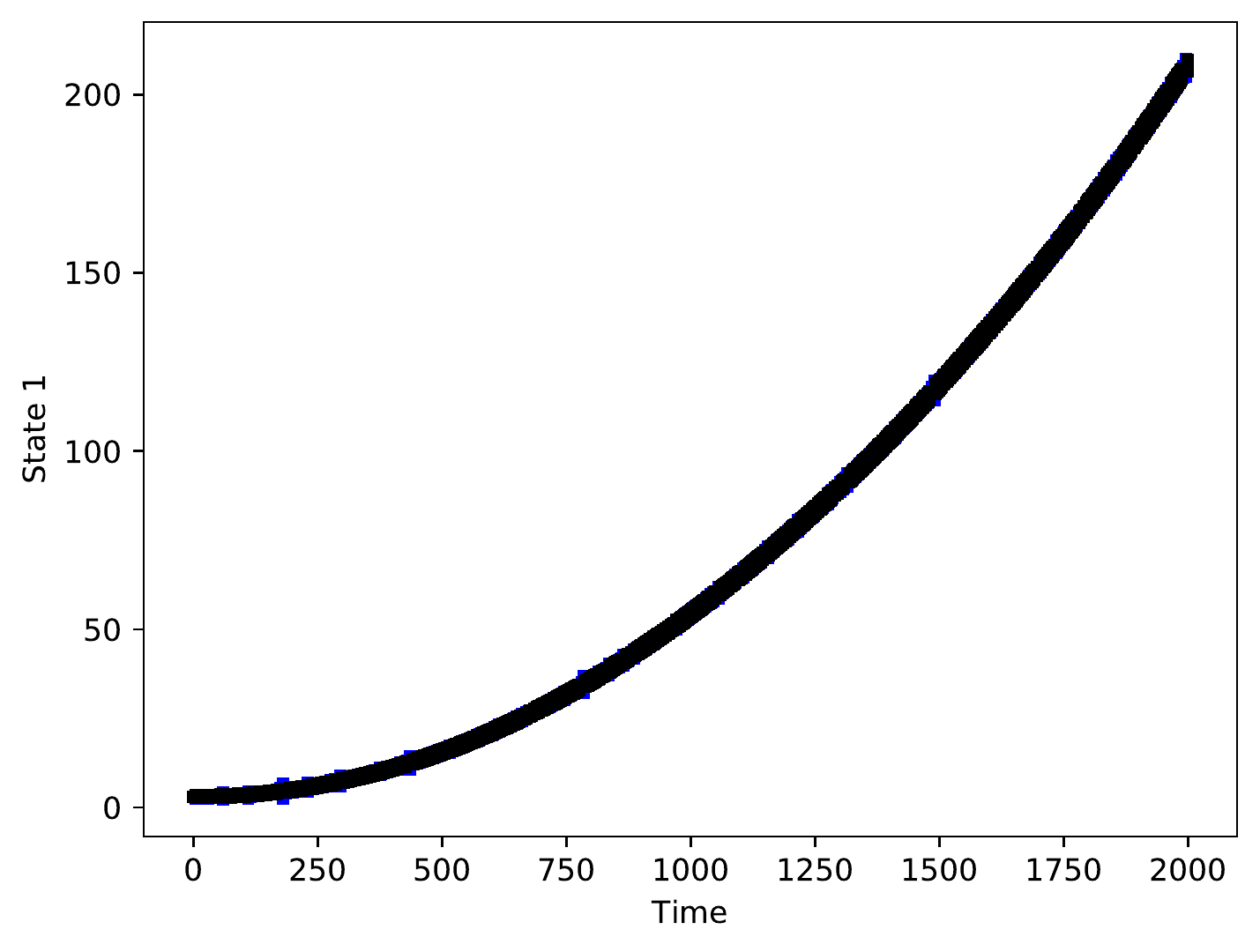}
        \caption{Frequent samples, high uncertainty}
        \label{fig:acc_fig4}
    \end{subfigure}
    
    \vfill
    
    \begin{subfigure}{.48\textwidth}
        \centering
        \includegraphics[width=\linewidth]{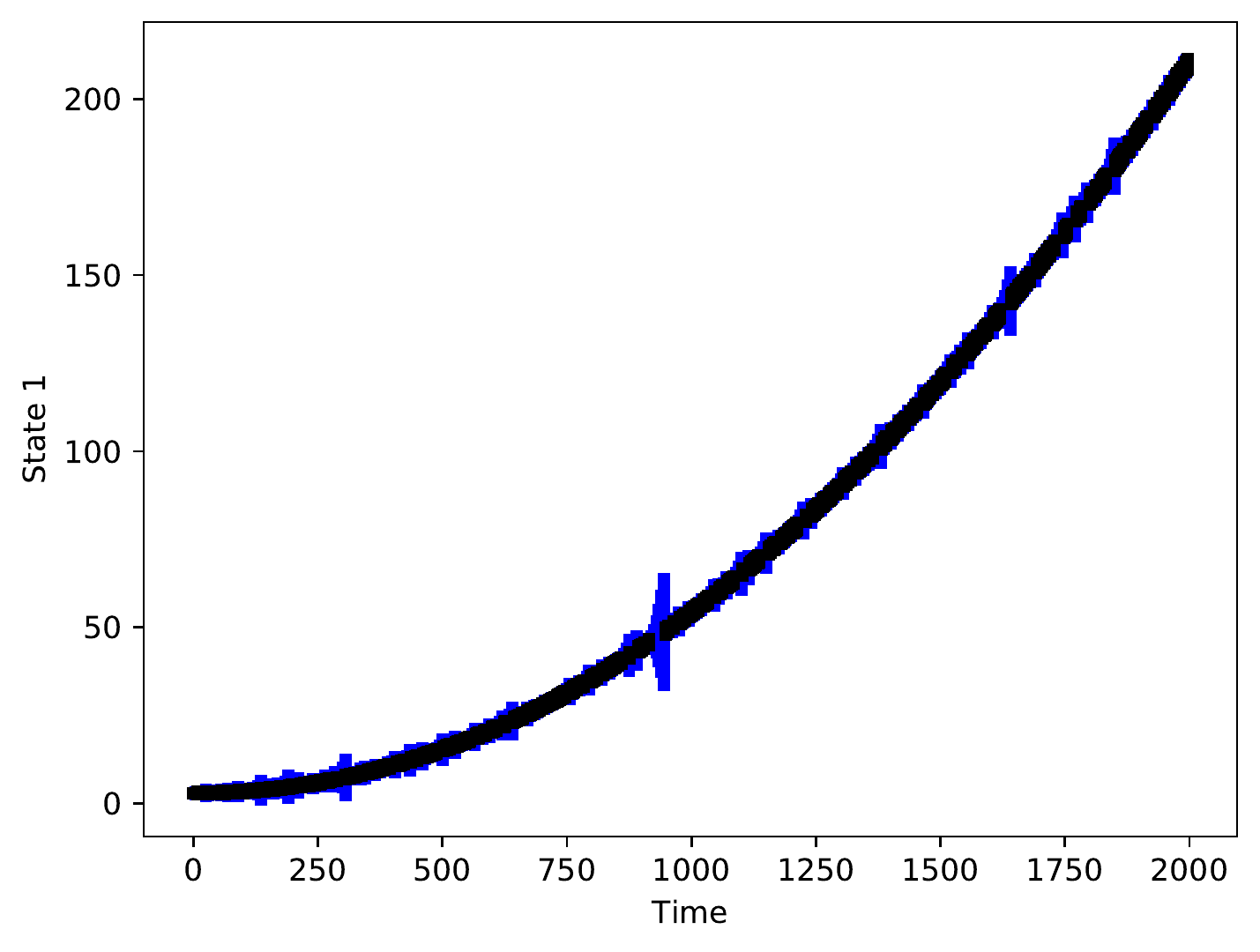}
        \caption{Sporadic samples, low uncertainty}
        \label{fig:acc_fig1}
    \end{subfigure}
    \hfill
    \begin{subfigure}{.48\textwidth}
        \centering
        \includegraphics[width=\linewidth]{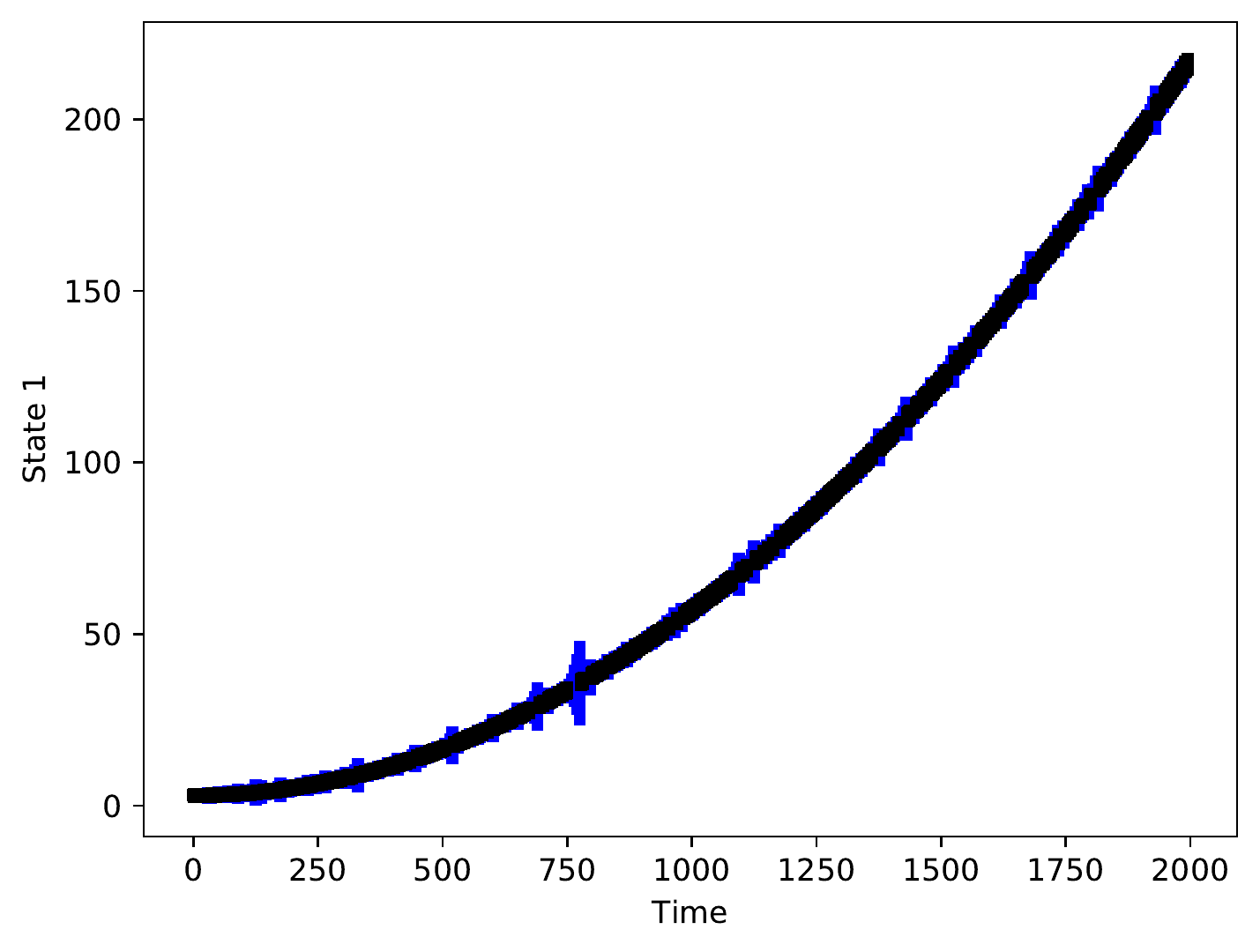}
        \caption{Sporadic samples, high uncertainty}
        \label{fig:acc_fig2}
    \end{subfigure}
    \caption{\textbf{\textit{Offline Monitoring (ACC).}} We plot the change in distance $h$ between the vehicles with time. The volume of the samples increases from left to right, and the probability of logging increases from bottom to top.} 
    \label{fig:offlineACC}
\end{figure*}

\begin{figure*}
    \begin{subfigure}{.49\textwidth}
        \centering
        \includegraphics[width=\linewidth]{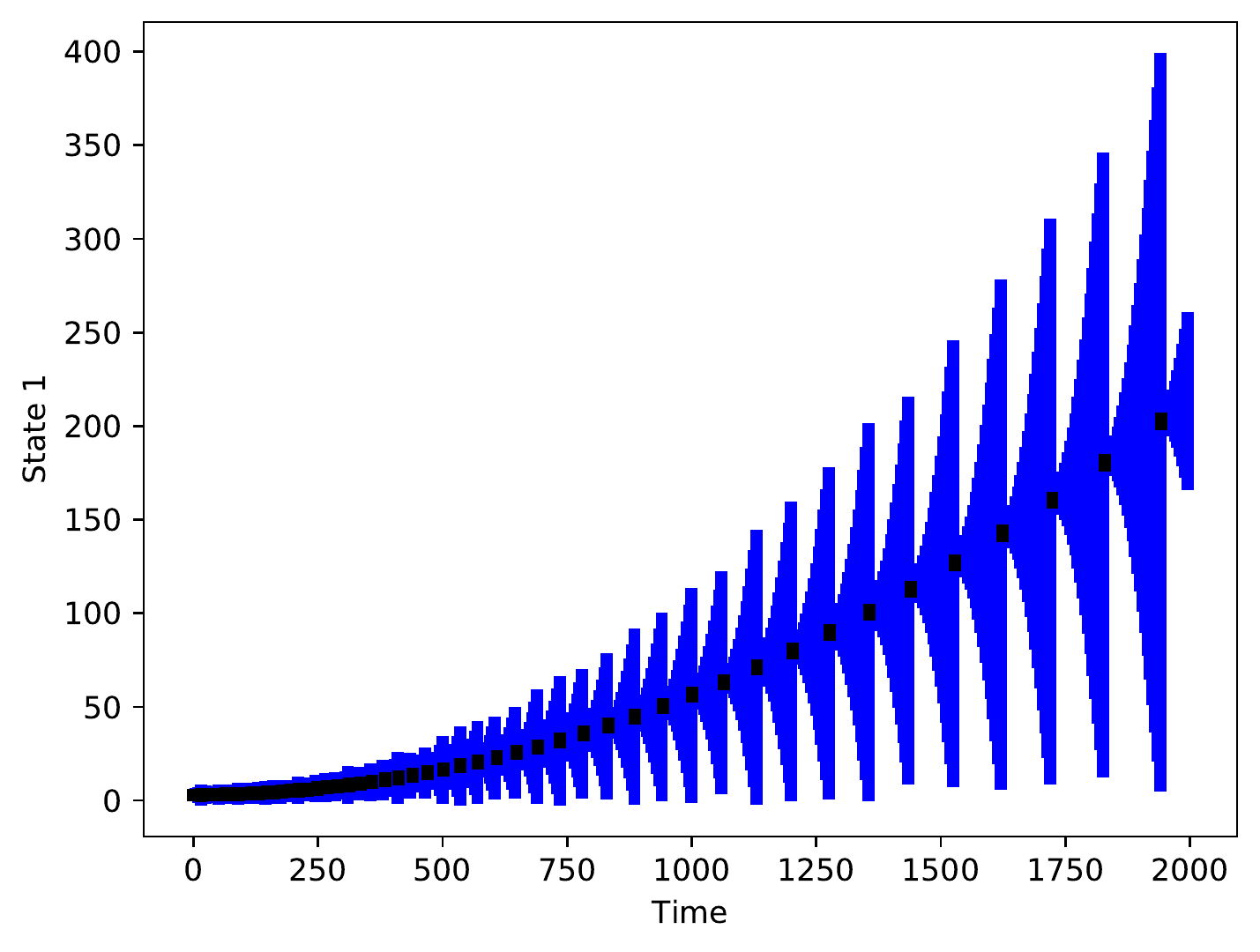}
        \caption{\textbf{Online monitoring}}
        \label{fig:acc_online}
    \end{subfigure}
    \hfill
    \begin{subfigure}{.49\textwidth}
        \centering
        \includegraphics[width=\linewidth]{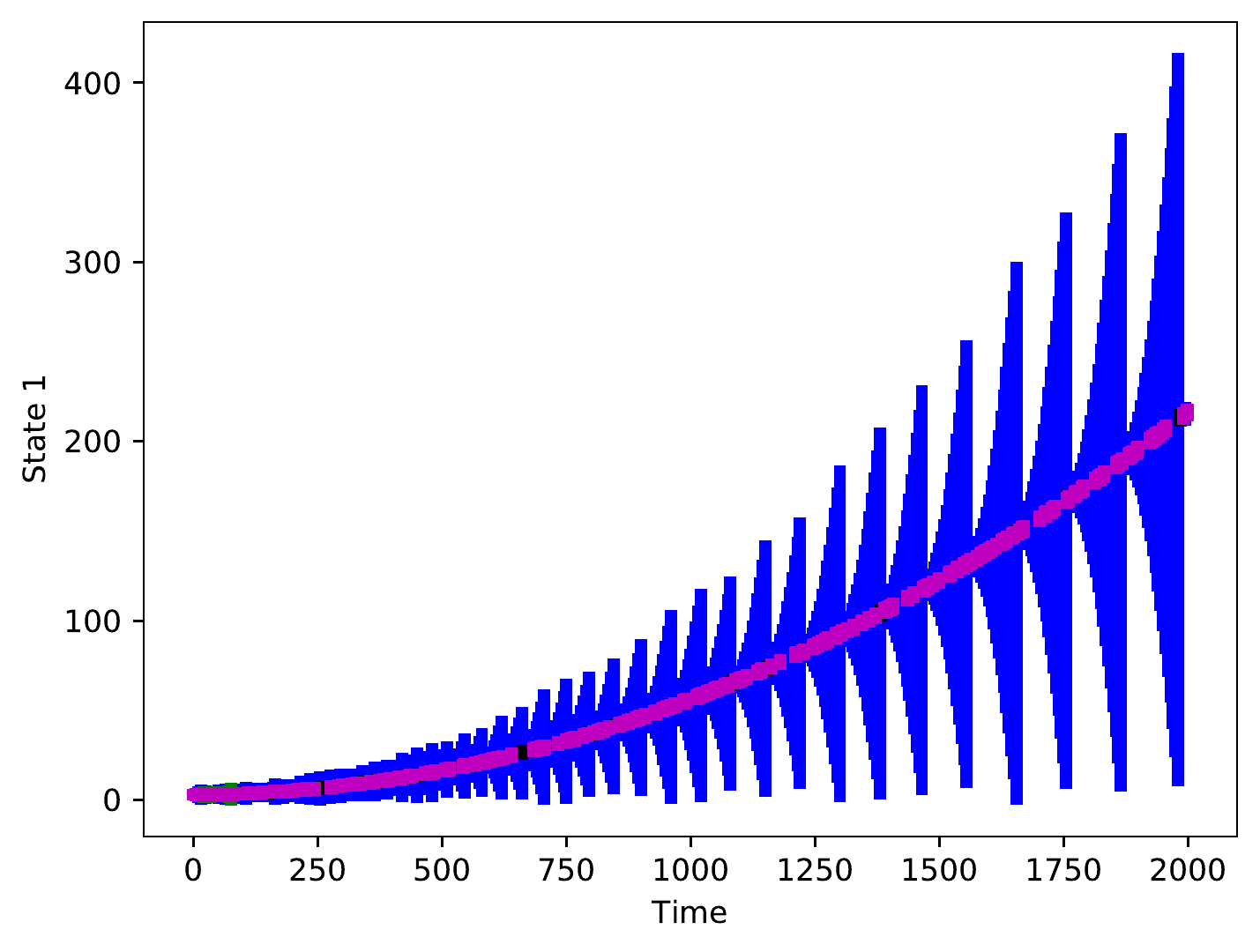}
        \caption{\textbf{Comparing online and offline monitoring}}
        \label{fig:acc_comp}
    \end{subfigure}
    \caption{\textbf{\textit{Online Monitoring (ACC).}} We plot the change in distance between two vehicle $h$ with time. The color coding is same as \cref{fig:online}. \textit{Online monitoring (\cref{fig:acc_online})}: We apply our online monitoring to the ACC model.
    \textit{Comparison (\cref{fig:acc_comp})}: We compare our online and offline algorithms.}
    \label{fig:onlineACC}
\end{figure*}

\subsubsection{Experiments}\label{ACC:experiments}

Next, we answer questions (1)-(4), using \cref{fig:offlineACC,fig:onlineACC}.
In \cref{fig:offlineACC}:
\begin{ienumerate}%
	\item the plots in the bottom row have logging probability of 20\,\%, and the plots in top row have a logging probability of 40\,\%;
	\item the plots in left column and the right column have been simulated with an initial set of $[\begin{smallmatrix}
[15,15.01] & [3,3.03] & [14.9,15]
\end{smallmatrix}]^\top
$ and
$
[\begin{smallmatrix}
[15,15.1] & [3,3.5] & [14.9,15.1]
\end{smallmatrix}]^\top
$
respectively.
\end{ienumerate}

\paragraph{Initial state}\label{ACC:initial}
In \cref{fig:online}, we simulated the trajectory with an initial set\linebreak[4]$[\begin{smallmatrix}
[15,15.01] & [3,3.03] & [14.9,15]
\end{smallmatrix}]^\top
$, $u \in [2,5]$.
These initial conditions are chosen such that the cars start from a safe specification.

\paragraph{Answer to Question~1}
	We answer this question by comparing two sets figures in the left column (\cref{fig:acc_fig1,fig:acc_fig3}) and the right column (\cref{fig:acc_fig2,fig:acc_fig4}) of \cref{fig:offlineACC}. \textit{For the left column, \ie{} with smaller sample size}: \cref{fig:acc_fig1} took 19.08\,s and concluded the system to be safe. This analysis in this plot invoked the refinement module of the offline algorithm. But increasing the probability of logging, \ie{} more number of samples, as in \cref{fig:acc_fig3}, resulted in not invoking the refinement module at all, thus taking 16.5\,s. \textit{For the right column, \ie{} with larger sample size}: The analysis is similar to that of the left column.
	\cref{fig:acc_fig2} invoked the refinement module several times, thus taking 20.84\,s, while \cref{fig:acc_fig4} took 17.5\,s, as it invoked the refinement module a smaller number of times.

\paragraph{Answer to Question~2}
We answer this question by comparing two sets figures in the top row (\cref{fig:acc_fig3,fig:acc_fig4}) and the bottom row (\cref{fig:acc_fig1,fig:acc_fig2}) of \cref{fig:offlineACC}. \textit{For the bottom row, \ie{} with smaller logging probability}: Comparing \cref{fig:acc_fig1} and \cref{fig:acc_fig2} shows that increasing sample volume results in invoking the refinement module more frequently.
    A very similar behavior is seen by comparing the top row (\ie{} with higher logging probability).

\paragraph{Answer to Question~3}
	Using our online algorithm, we were able to prove safety of the system in 104.58\,s. The online algorithm triggered the logging system to generate samples for 53 time steps---this is less than 3\,\% of total time steps. This is shown in \cref{fig:acc_online}.

\paragraph{Answer to Question~4}
	We compare our offline and online algorithm, for $2\,000$ time steps, on the same trajectory.
	The result is given in \cref{fig:acc_comp}. Note that, using our online algorithm, we were able to prove safety of the system in~124.46\,s. The online algorithm triggered the logging system to generate samples only 50~times.
	In contrast, the offline algorithm, with a log size of~281 (14\,\% logging probability) took 28.54\,s to infer that the system is safe.
\subsection{Third benchmark: Aircraft Orbiting}\label{subsection:aircraft}

Next, we apply our {offline} algorithm to an aircraft orbiting case study.
Unlike the former two case studies, we consider in this third benchmark not only uncertain valuations, but also uncertain timestamps.
That is, logs for this benchmark are uncertain logs (as in \cref{definition:log}).

\subsubsection{System description}
\begin{figure}[tb]
  \centering
 \includegraphics[width=.6\textwidth]{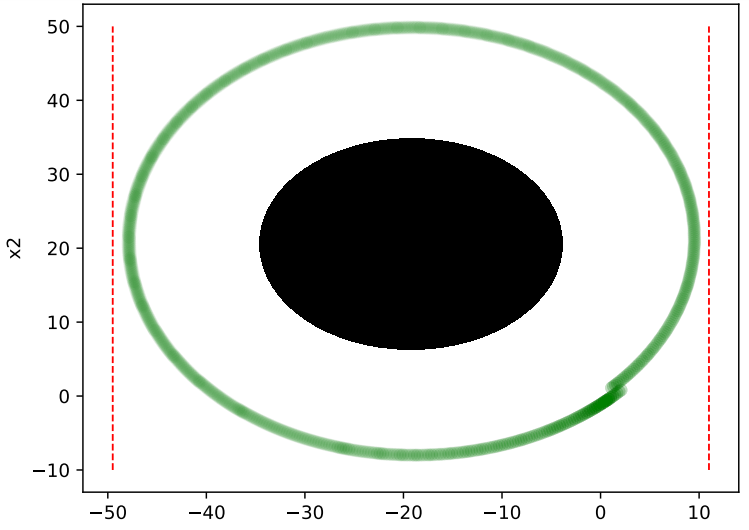}
 \caption{\textit{Planned behavior of the aircraft}: The axis of the plot denotes the position of the aircraft in $(x,y)$  plane (with a fixed altitude). The ideal (planned) path of the aircraft, orbiting the object (black), is shown in green. The red dashed lines indicate the safety constraint of the aircraft---these lines must not be crossed at any time step.}
 \label{fig:behAirctaft}
\end{figure}

In this case study, an aircraft is orbiting an object to gather information about the object---depicted in \cref{fig:behAirctaft}.
The aircraft orbits the object by controlling its angular velocity.
Since the aircraft operates in an uncertain environment, the angular velocity of the aircraft can have added noise.
That is, the model of the aircraft used in this case study can have uncertainties in its angular velocity.
The axis of \cref{fig:behAirctaft} denotes the position of the aircraft
in $(x,y)$ plane (we assume a fixed altitude).
The ideal (planned) path of the aircraft,
orbiting the object (black), is shown in green.
The aircraft must always stay sufficiently close to the object for the data collected about it to be considered reliable.
One such bound is shown in the plot with red dashed lines---the aircraft must not cross these lines at any time step.
The aircraft transmits its positional data (the $x, y$ values) to the local station at
aperiodic intervals. Since the data is transmitted over a shared, long-distance network, a delay is experienced when the data reaches the local station.
In other words, the timestamps of the log (behavioral data) can have added noise.
Additionally, the behavioral data can also have added noise due to sensor uncertainties.
Therefore, this case study is an ideal candidate to demonstrate the applicability of our offline monitoring approach with added noise in timestamps (\cref{ss:offline-uncertain}).

\paragraph{Model}
The model as in \cite{lal,PC09} has the following state variables:
\begin{ienumerate}%
	\item position of the aircraft in two dimensional plane $(x_1,x_2)$,
	\item and velocity $(d_1,d_2)$.
\end{ienumerate}%
The state space of the system is given in~\cite{lal}.
The set of state variables of this system is
$
[\begin{smallmatrix}
x_1 & x_2 & d_1 & d_2
\end{smallmatrix}]^\top
$.

\paragraph{Model parameters}
The model is dependent on the angular velocity $\omega$ of the aircraft. We assume $\omega \in 1.5 \pm 1\%$.

\paragraph{Safety}
The behavior of the aircraft is considered safe, under the presence of model uncertainties, if $x \in [-49.5, 11]$. The safety condition is defined as the state in which the radius of the aircraft orbiting the object remains within a reasonable range. This constraint is necessary to ensure that the sensors can effectively gather data on the object without encountering difficulties caused by an excessively large radius.

\subsubsection{Experiments}\label{paragraph:aircraft:experiments}
Recall that in the first two case studies (Anesthesia and ACC), we selected logging probabilities of~20\,\% and~40\,\% to represent sporadic and frequent logging, respectively. However, in this particular case study, we have chosen a logging probability of~5\,\% for sporadic logging and~10\,\% for frequent logging.
This choice was made because this case study differs significantly from the two others: through our empirical observations, we found that the former logging probabilities (20\,\% and 40\,\%) never provided an opportunity to observe any unsafe behavior in this case study.
The computed reachable sets, which represent the possible system states between two logs, were so tightly over-approximated that they never intersected with the unsafe region. 
Consequently, we had to decrease the logging probability in order to expand the reachable sets and increase the likelihood of identifying potential unsafe conditions.
To incorporate this different notion of sporadicity in our experiments, we have empirically chosen logging probabilities of~5\,\% for sporadic logging and~10\,\% for frequent logging.

\begin{figure*}
    \begin{subfigure}{.48\textwidth}
        \centering
        \includegraphics[width=\linewidth]{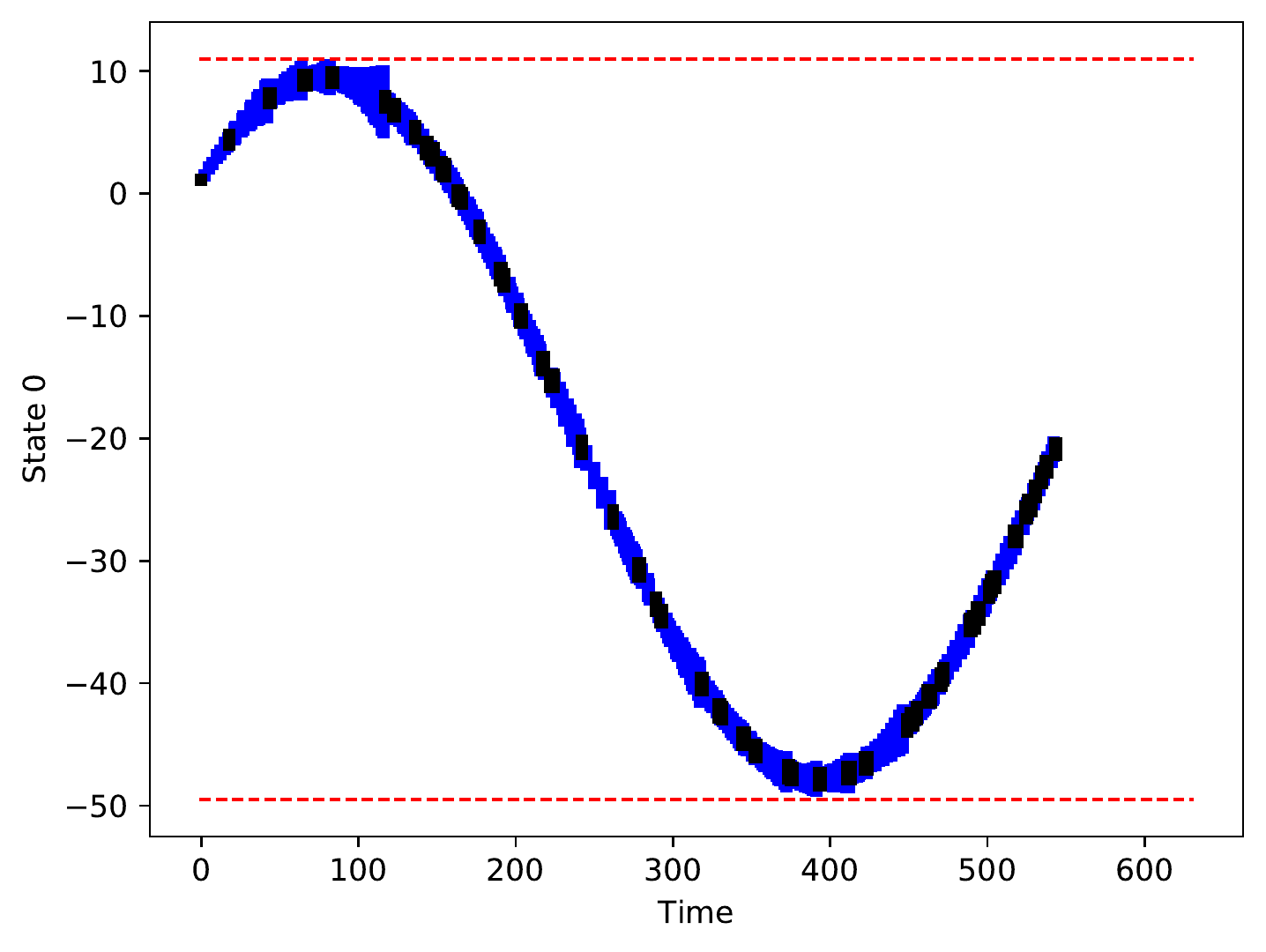}
        \caption{Frequent samples, low uncertainty}
        \label{fig:air_fig3}
    \end{subfigure}
    \hfill
    \begin{subfigure}{.48\textwidth}
        \centering
        \includegraphics[width=\linewidth]{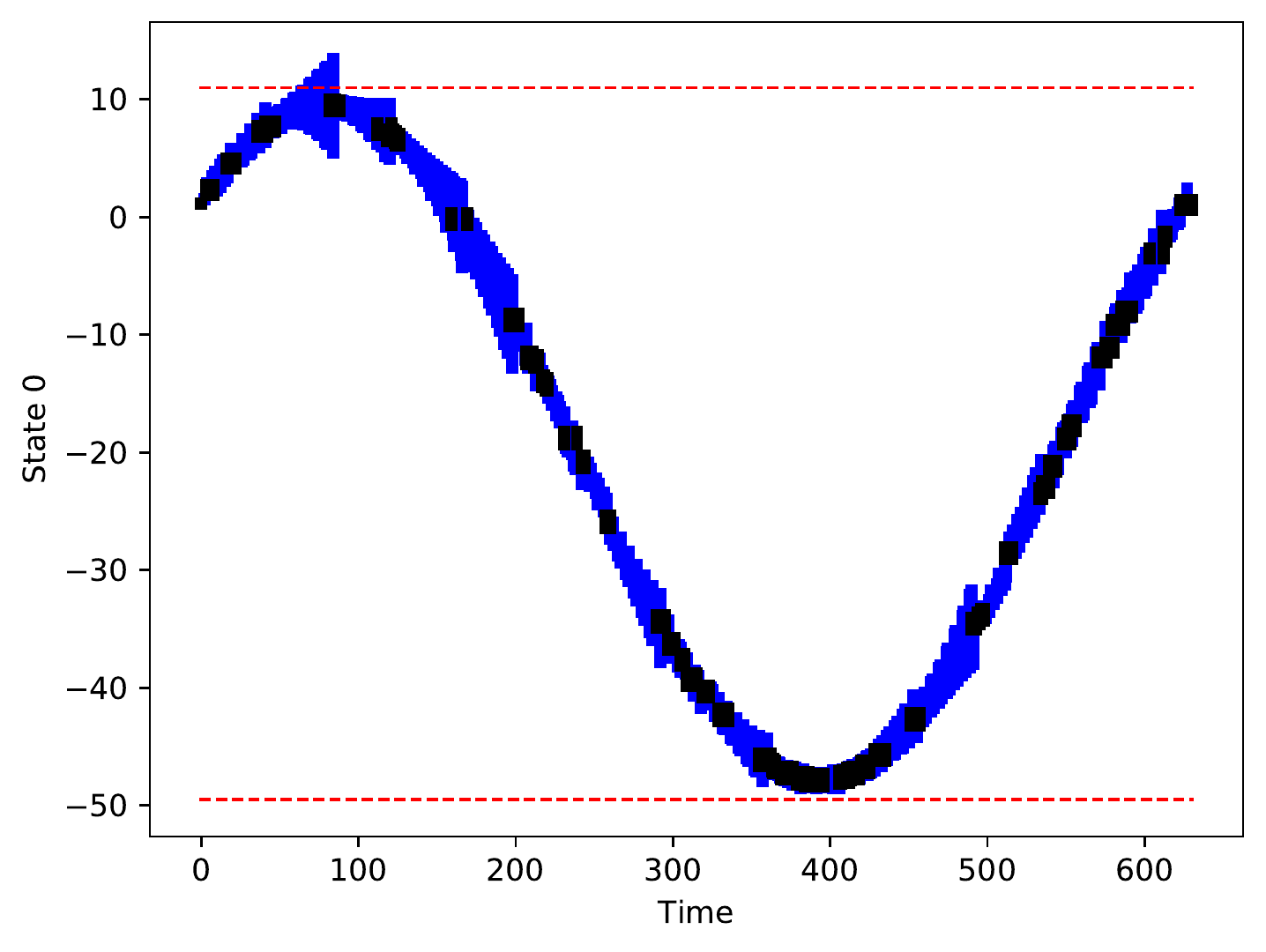}
        \caption{Frequent samples, high uncertainty}
        \label{fig:air_fig4}
    \end{subfigure}
    
    \vfill
    
    \begin{subfigure}{.48\textwidth}
        \centering
        \includegraphics[width=\linewidth]{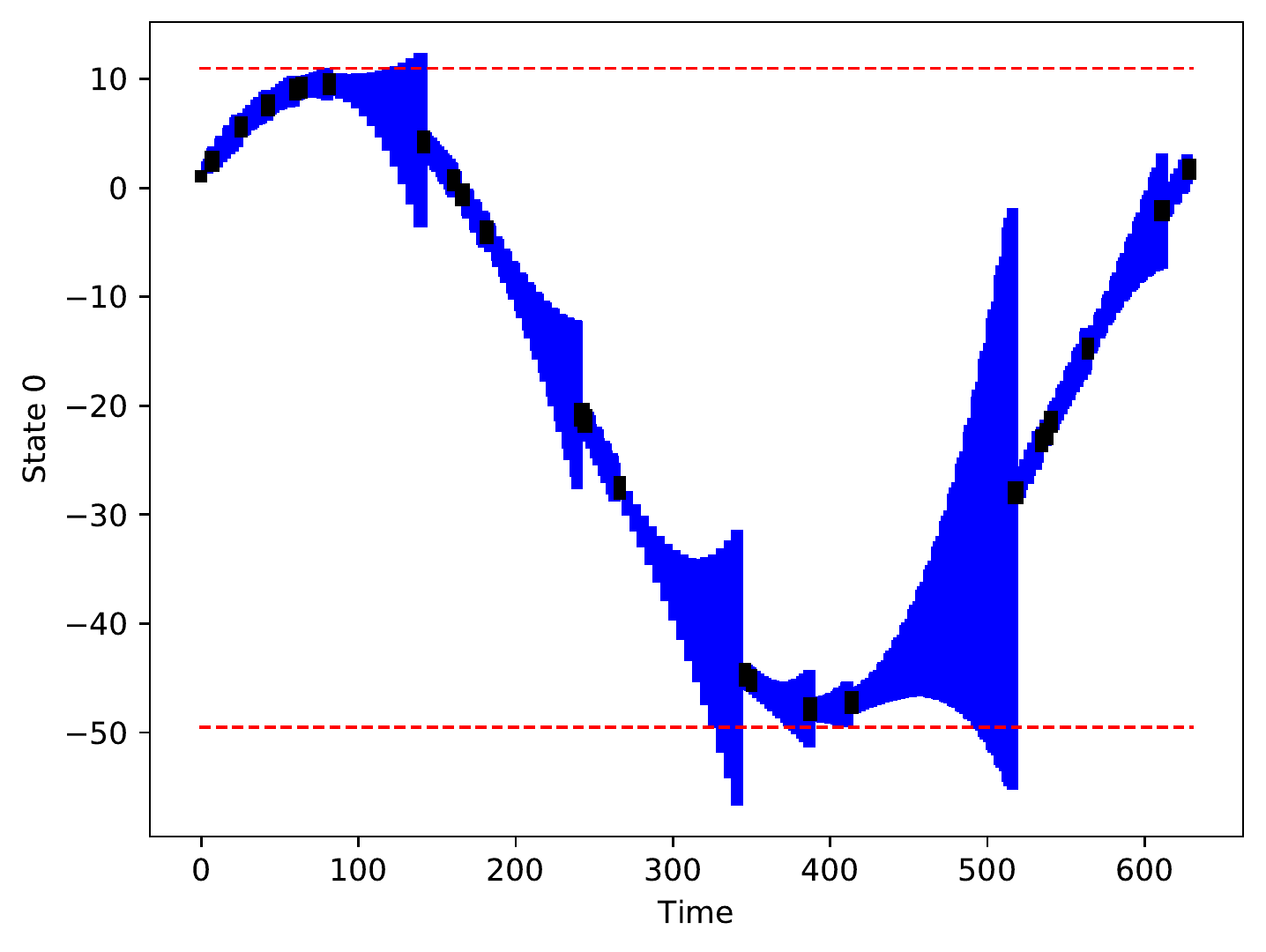}
        \caption{Sporadic samples, low uncertainty}
        \label{fig:air_fig1}
    \end{subfigure}
    \hfill
    \begin{subfigure}{.48\textwidth}
        \centering
        \includegraphics[width=\linewidth]{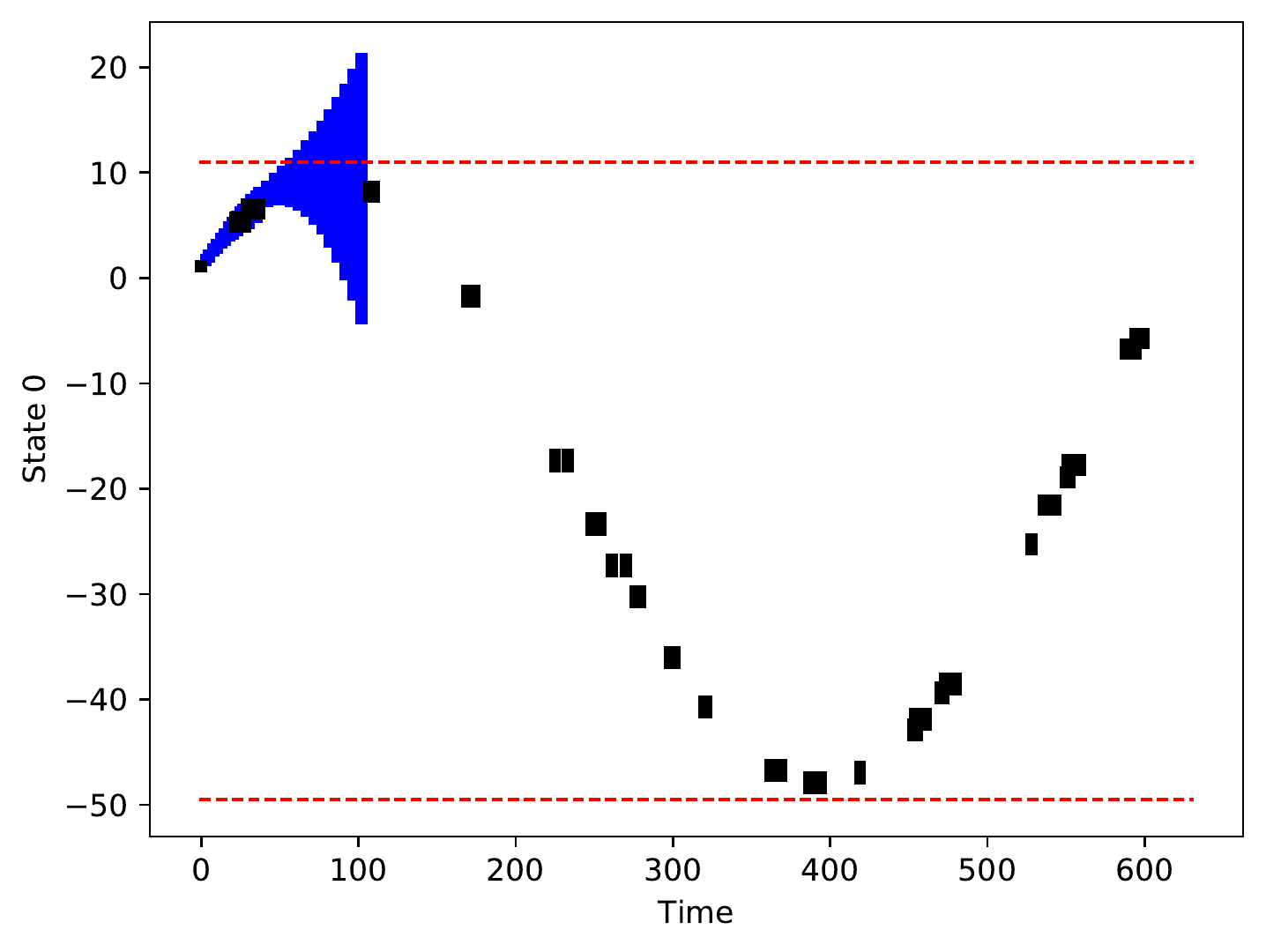}
        \caption{Sporadic samples, high uncertainty}
        \label{fig:air_fig2}
    \end{subfigure}
    \caption{\textit{Offline monitoring (Aircraft Orbiting)}:  We plot the position of the aircraft, along $x$ axis, with time. The timing delay of the samples increases from left to right, and the probability of logging increases from bottom to top.
 {The color coding is same as \cref{fig:offline}.}}
    \label{fig:offlineAircraft}
\end{figure*}

\begin{figure}
    \centering
    \includegraphics[width=0.48\textwidth]{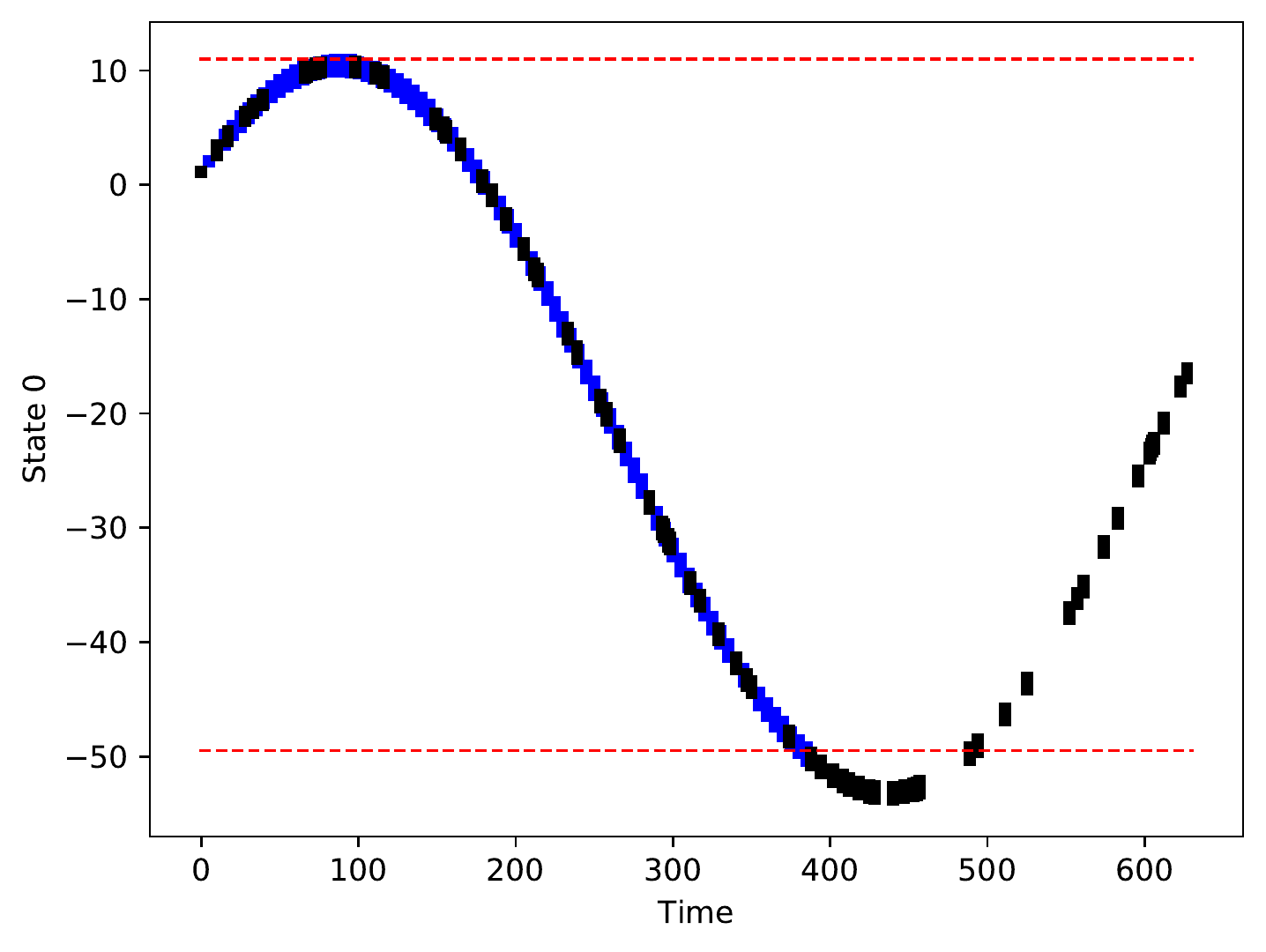}
    \caption{\textit{Offline monitoring with unsafe samples (Aircraft Orbiting):} We plot the position of the aircraft, along $x$ axis, with time. We perform offline monitoring on a log containing unsafe samples.
 {The color coding is same as \cref{fig:offline}.}}
\label{fig:offlineAircraftUnsafeLog}
\end{figure}

\begin{figure*}  
    \begin{subfigure}{.48\textwidth}
        \centering
        \includegraphics[width=\linewidth]{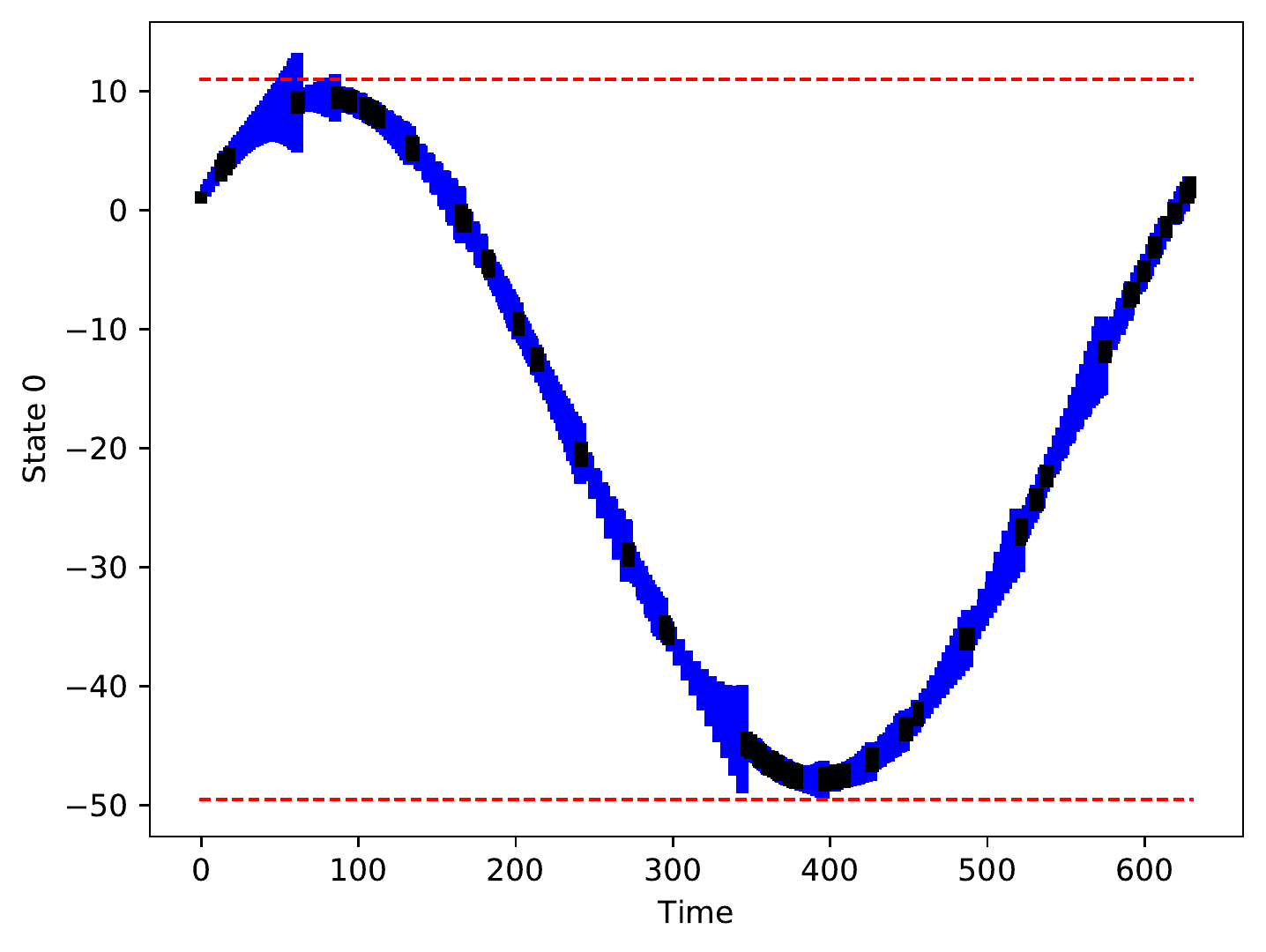}
        \caption{Time delay of 2 units.}
        \label{fig:air_fig5}
    \end{subfigure}
    \hfill
    \begin{subfigure}{.48\textwidth}
        \centering
        \includegraphics[width=\linewidth]{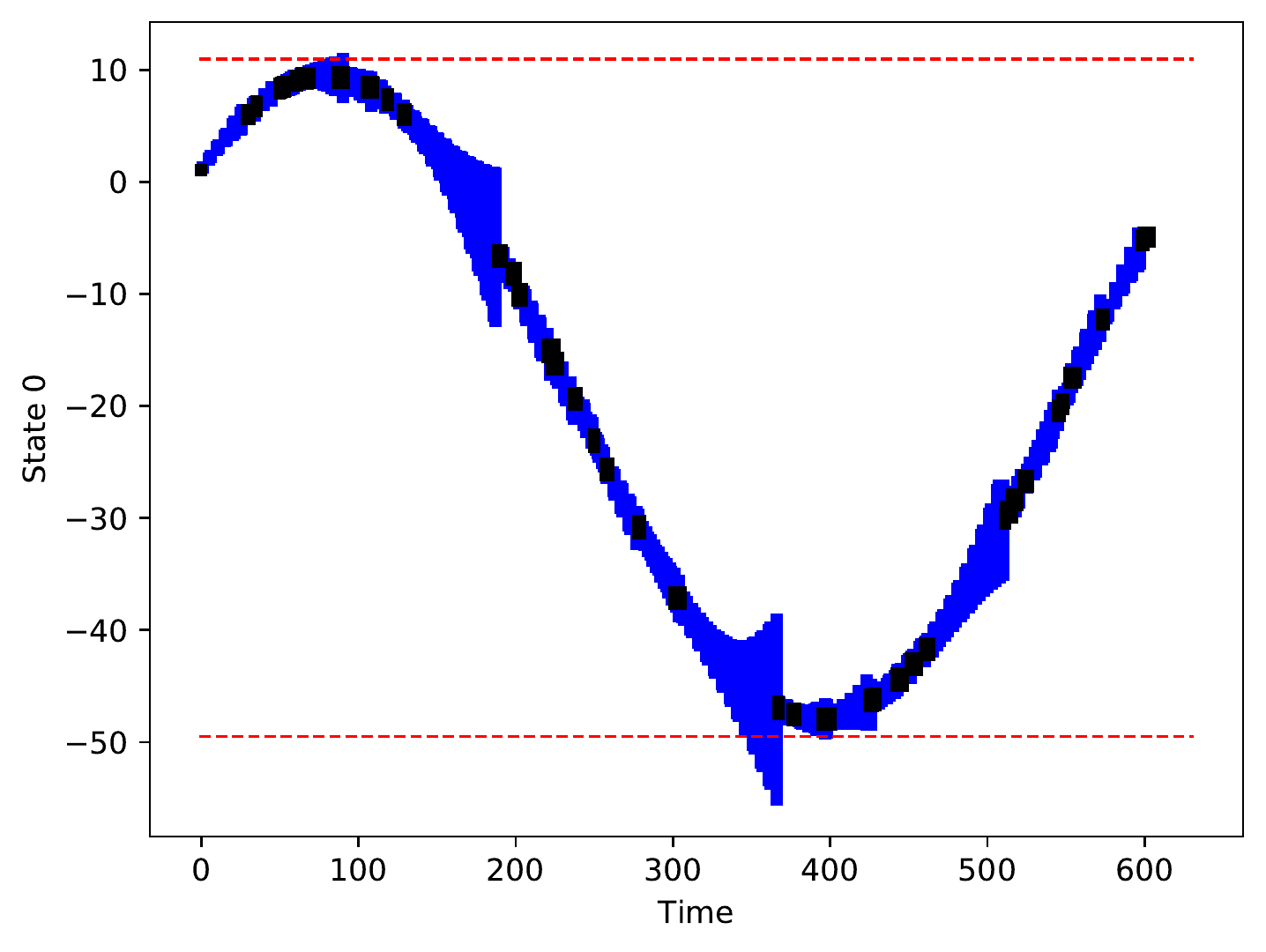}
        \caption{Time delay of 6 units.}
        \label{fig:air_fig6}
    \end{subfigure}
    \\
    \centering
    \begin{subfigure}{.48\textwidth}
        \centering
        \includegraphics[width=\linewidth]{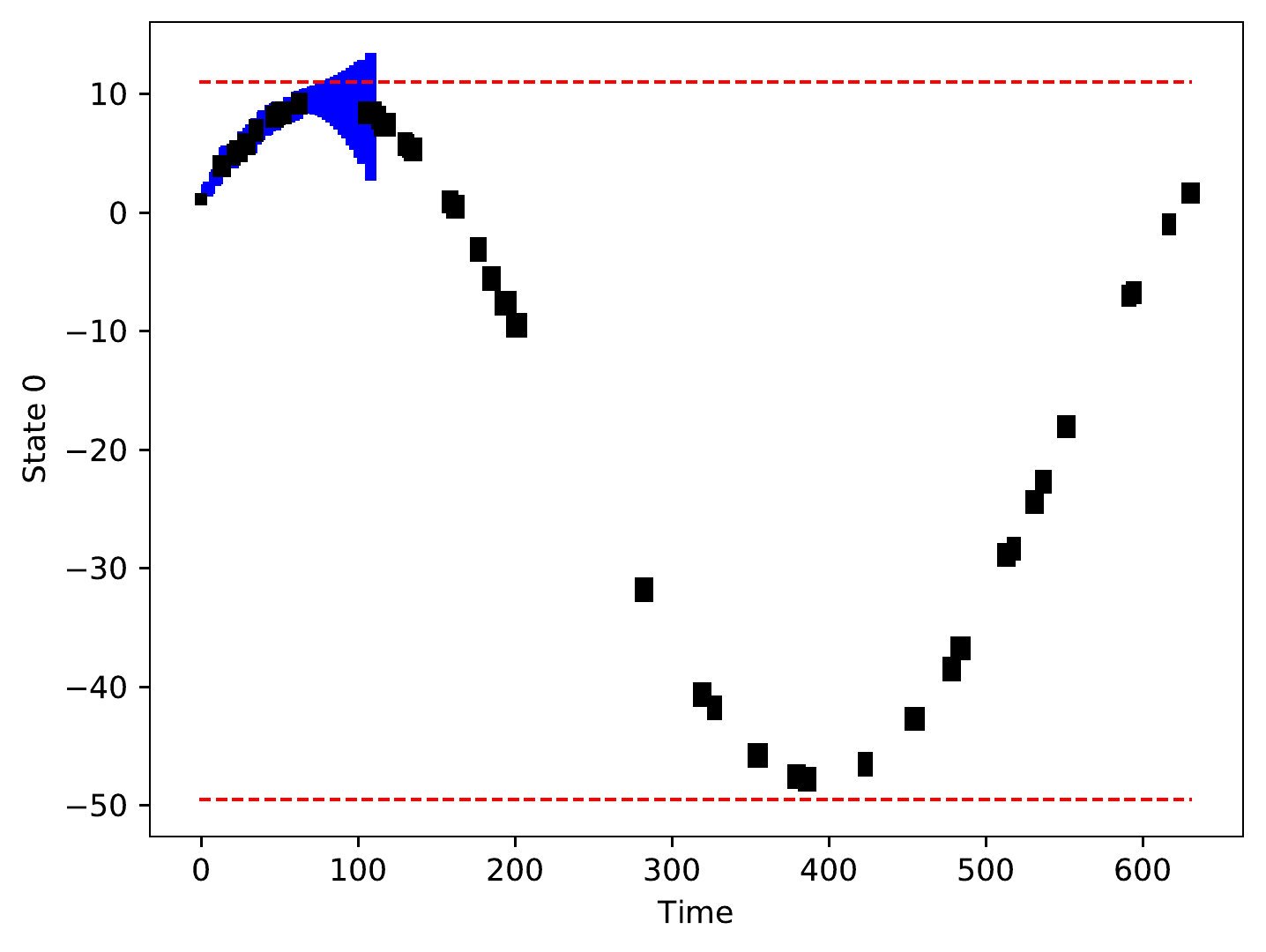}
        \caption{Time delay of 8 units.}
        \label{fig:air_fig7}
    \end{subfigure}
    \caption{\textit{Effect of timing delay of samples on offline monitoring (Aircraft Orbiting)}:  We plot the position of the aircraft, along $x$ axis, with time. The timing delay of the samples increases from left to right, while the probability of logging remains constant at 7\,\% across all plots.
 {The color coding is same as \cref{fig:offline}.}}
    \label{fig:offlineAircraftUT}
\end{figure*}

Additionally, to further evaluate the impact of logging probabilities and the uncertainty in sample timestamps, we also conducted experiments using an intermediate logging probability of~7\,\% with three different choices of timestamp uncertainties.
In \cref{fig:offlineAircraft}, the plots in the bottom and top rows have a logging probability of 5\,\% (sporadic sampling) and 10\,\% (frequent sampling) respectively, and the plots in the left column and the right column have a timing delay of 2~units and 10~units respectively.
\cref{fig:offlineAircraftUnsafeLog} contains unsafe samples and has a logging probability of 10\,\% (with no time delay), which distinguishes it from the plots in \cref{fig:offlineAircraft}. In \cref{fig:offlineAircraftUT}, the plots within it have a logging probability of 7\,\%, while the time delay progressively increases from left to right, with time delays of 2, 6, and 8 units respectively.

\paragraph{Initial state}\label{aircraft:initial}
All the plots in \cref{fig:offlineAircraft,fig:offlineAircraftUnsafeLog,fig:offlineAircraftUT} have been generated with an initial set of
$[\begin{smallmatrix}
[1.1,1.11] & [1.1,1.11] & [20,20.1] & [20,20.1]
\end{smallmatrix}]^\top
$.
The initial set is chosen as per the orbiting path of the aircraft, and such that it remains sufficiently close to the object.

\medskip

We now answer Questions~(1) and~(5) using \cref{fig:offlineAircraft,fig:offlineAircraftUnsafeLog,fig:offlineAircraftUT}.

\paragraph{Answer to Question~1} The observations \emph{vis-à-vis} this question are very similar to previous two case studies.
We answer this question by comparing two sets figures in the left column (\cref{fig:air_fig1,fig:air_fig3}) and the right column (\cref{fig:air_fig2,fig:air_fig4}) of \cref{fig:offlineAircraft}. \textit{For the left column, \ie{} with smaller timing delay}: \cref{fig:air_fig1} took  111.6\,s and concluded the system to be safe. The analysis in this plot invoked the refinement module of the offline algorithm. But increasing the probability of logging, \ie{} more number of samples, as in \cref{fig:air_fig3}, resulted in not invoking the refinement module at all, thus taking 31.61\,s. \textit{For the right column, \ie{} with larger timing delay}: this analysis, as shown in \cref{fig:air_fig2}, took 2.33\,s to complete, and concluded the system behavior to be unsafe. The behavior of the system, shown in \cref{fig:air_fig4} with 10\,\% probability of logging, results in inferring the behavior of the system as safe, by invoking the refinement module. Overall, this analysis, as shown in \cref{fig:air_fig4}, took 48.39\,s to complete, and concluded the system behavior to be safe.
Although none of the plots depicted in \cref{fig:offlineAircraft} exhibited any unsafe samples, the inclusion of \cref{fig:offlineAircraftUnsafeLog} demonstrates how our offline monitoring approach can trivially identify such unsafe samples and label the system's behavior as unsafe.
\cref{fig:offlineAircraftUnsafeLog} accurately labeled the system's behavior as unsafe within a detection time of 4.98\,s upon encountering the unsafe sample.

\paragraph{Answer to Question~5} We answer this question by comparing two sets figures in the top row (\cref{fig:air_fig3,fig:air_fig4}) and the bottom row (\cref{fig:air_fig1,fig:air_fig2}) of \cref{fig:offlineAircraft}. \textit{For the bottom row, \ie{} with smaller logging probability}: Increasing the timing delay of the samples results in inferring the behavior from safe (\cref{fig:air_fig1}) to unsafe (\cref{fig:air_fig2}), as per the offline monitoring algorithm. \textit{For the top row, \ie{} with higher logging probability}: Increasing the timing delay the samples results in not invoking the refinement module (\cref{fig:air_fig3}) to invoking the refinement module several times (\cref{fig:air_fig4}), as per the offline monitoring algorithm. 
The same observation is further reinforced by \cref{fig:offlineAircraftUT}. Specifically, \cref{fig:air_fig5,fig:air_fig6,fig:air_fig7} exhibit timing delays of 2, 6, and 8~units respectively, while maintaining the same logging probability across all plots. Notably, we observe that when the timing delays were 2 and 6~units (\cref{fig:air_fig5,fig:air_fig6} respectively), our offline monitor inferred the behavior to be safe in 48.38\,s and 56.48\,s respectively.
It is worth mentioning that the time required by the offline monitor increases as the time delay of the sample increases.
However, in contrast, \cref{fig:air_fig7} illustrates an instance where the offline monitor inferred
	the system behavior as unsafe, accomplishing this in~4.97\,s. Consequently, as also observed in \cref{fig:offlineAircraft}, \cref{fig:offlineAircraftUT} demonstrates that an increase in the time delay of the samples can potentially lead to false alarms by the offline monitor, along with an increase in computation time.

Let us now discuss the reasons behind selecting a logging probability of~7\,\% to generate \cref{fig:offlineAircraftUT} (while logging probabilities of 5\,\% and~10\,\% were chosen for \cref{fig:offlineAircraft}).
A higher logging probability results in fewer false alarms, while a higher timing delay also reduces false alarms. Consequently, with a logging probability of~5\,\% and a timing delay of 10~units, the system is inferred as unsafe. However, with a logging probability of~10\,\%, a timing delay of 10~units was successfully verified.
In contrast, for \cref{fig:offlineAircraftUT}, we chose a logging probability of~7\,\% to empirically determine the tolerable amount of timing delay. Through our experiments, we discovered that with a logging probability of~7\,\%, the system was able to tolerate timing delays of 2 and 6~units. However, at a timing delay of 8~units, the system was deemed unsafe.

\subsection{General observations} \label{subsec:genObs}
In the following, we provide general answers to questions (1)-(5) based on our observations from our three case studies.
\paragraph{Answer to Question~1}
    Increasing the probability of logging reduces the chances of inclusion of spurious behaviors due to over-approximate reachable set computation over longer time horizon. Therefore, it has a reduced chance of spuriously inferring the system unsafe, also fewer chance of invoking the refinement module (as there are less spurious behaviors).

\paragraph{Answer to Question~2}
	Increasing the size of samples (due to uncertainties or inherent nature of the system) results in increasing chances of invoking the refinement module more frequently.
    It also increases the chance of (wrongly) inferring the system to be unsafe, as the refinement module can in itself add to the overapproximation.

\paragraph{Answer to Question~3}
	We observed that our online algorithm is able to prove the system's safety very efficiently with very few samples.

\paragraph{Answer to Question~4}
	We observed that for a given random log, the offline algorithm was unable to prove safety of the system, whereas our online algorithm was able to prove safety of the system, using fewer samples, by intelligently sampling the system only when needed.
	We also note that, though here we just demonstrated the result for one random log, but our internal experiments showed that the online algorithm always needed fewer samples to prove safety---which is unsurprising, as it is designed to sample the system only when needed.
	This can also result in energy saving, as sampling usually requires energy and bandwidth.
	
	\paragraph{Answer to Question~5}
	Increasing the timing delay of samples results in increasing chances of invoking the refinement module more frequently.
    It also increases the chance of (wrongly) inferring the system to be unsafe, as the refinement module can in itself add to the overapproximation. Further, it increase in computation time---as it requires exploring all possible combinations of timing delays.

\paragraph{Discussion: Reachable sets computation using \flowstar{}}
As uncertain linear dynamical systems are a special type of non-linear systems, \flowstar{}~\cite{CAS13} would have been a natural candidate to benchmark our offline and online monitoring implementation by comparing various methods to compute $\reach{\cdot}$. However, we ran into the following issues:
\begin{ienumerate}%
	\item To the best of our understanding, \flowstar{} expects the model of the continuous dynamics to be given as input, along with a discretization parameter. Therefore, trying to encode the time-varying uncertainties in the system as state variables will lead to discretization of the variables encoding uncertainties; such discretization leads to undesired behavior, as those uncertain variables will fail to capture the actual range of values that are possible at any time step.
	\item However, \flowstar{} does allow time varying uncertainties, but only additive\footnote{See example at \url{https://flowstar.org/benchmarks/2-dimensional-ltv-system/}}.
	Unfortunately, {all our case studies require} \emph{multiplicative} uncertainties.
\end{ienumerate}%
Still, we believe \flowstar{} could be compared with our implementation when the bounding model has a simpler dynamics than our uncertain linear dynamical systems.

\section{Conclusion}\label{section:conclusion}
\subsection{Summary}

We presented a new approach for monitoring cyber-physical systems against safety specifications, using the additional knowledge of an over-approximation of the system expressed using an uncertain linear dynamical system.
Our approach assumes as first input a log with scattered timestamps (either exact or given in the form of intervals) and uncertain variable samplings (in the form of zonotopes), and as second input an over-approximated model, bounding the possible behaviors.
The over-approximation is modeled by \emph{uncertainty} in the variables of the dynamics.

In the offline setting, we are thus able to detect possible violations of safety properties, by extrapolating the known samples with the over-approximated dynamics, and if needed using a second reachability analysis to check whether the next sample is ``compatible'' with the possible unsafe behavior, \ie{} can be reached from the unsafe zone.
In the online setting, we are capable of \emph{decreasing} the number of samples, triggering a sample only when there might be a safety violation in a near future, based on the latest known sample and on the over-approximated model dynamics---increasing the energetic efficiency.

Our methods are sound in the sense that an absence of detection of violation by our method indeed guarantees the absence of an actual violation at any discrete time step.
In the online method, provided the samples are accurate, our method is in addition complete, \ie{} the method outputs \emph{safe} iff the actual system is safe at all discrete time steps.
Put it differently, we guarantee that \emph{not} triggering a sample at some time steps is harmless and will not lead to missing a safety violation.

\subsection{Future works}\label{ss:future}
\paragraph{Bounding model}
The presence of an over-approximated model makes sense, as proposed in~\cite{WAH22TCPS}; in our setting of an over-approximated model given by an uncertain linear dynamical system, some formal guarantees that this model indeed represents an over-approximation of the actual system remain to be exhibited.

In addition, the assumption of the presence of an over-approximated model is central to our work, and we used it in all our experiments, in the sense that the logs were indeed instances of the over-approximated model.
However, an interesting future work will be to partially lift this assumption, by allowing the log to (temporarily, locally) differ from the over-approximated model{, allowing for more freedom}.
In that case, a special care must be made on the approach's soundness.

\paragraph{Enumeration of time steps}
A possible threat to validity remains the \emph{enumeration} of time steps in both our algorithms (\cref{algo:offline:forall} in \cref{algo:offline} and \cref{algo:online:for} in \cref{algo:online}), which could slow down the analysis for very sparse logs---even though this did not seem critical in our experiments.
It is worth recalling that the reachable set computation method from~\cite{ghosh1} that we use in this paper scales very well when computing reachable sets for smaller time steps.
Consequently, if the time gap between two samples in a log is not very large (\eg{} an order of 500~steps), this technique computes reachable sets very quickly.
Further, if the time gaps become large, one can use the interval-based or the zonotope-based reduction methods proposed in~\cite[Section~5.2]{ghosh1} to improve the scalability of the reachable set computation.

In addition, using skipping methods (as in, \eg{} \cite{WHS17}) may help improving the efficiency of our approach.

\begin{figure}[tb]
	{\centering
	\begin{subfigure}[c]{.48\textwidth}
		\centering
		\begin{tikzpicture}[shorten >=1pt, scale=.6, yscale=.6, xscale=0.7, auto]

			\draw[->] (0, 0) --++ (0, 5.0) node[anchor=north east]{$x$};
			\draw[->] (0, 0) --++ (8.0, 0) node[anchor=north]{$t$};
			\draw[dashed, color=red, semithick] (0, 4) --++ (8.0, 0);

			\fill[sample] (0, 1.8) coordinate (s1) circle[];
			\fill[sample] (0.5, 1.4) coordinate (s2) circle[];
			\fill[sample] (1, 1.9) coordinate (s3) circle[];
			\fill[sample] (1.5, 2.1) coordinate (s4) circle[];
			\fill[sample] (2, 2) coordinate (s5) circle[];
			\fill[sample] (2.5, 1.9) coordinate (s6) circle[];
			\fill[sample] (3, 1.4) coordinate (s7) circle[];
			\fill[sample] (3.5, 1.2) coordinate (s8) circle[];
			\fill[sample] (4, 0.8) coordinate (s9) circle[];
			\fill[sample] (4.5, 1.6) coordinate (s10) circle[];
			\fill[sample] (5, 2.5) coordinate (s11) circle[];
			\fill[sample] (5.5, 2.8) coordinate (s12) circle[];
			\fill[sample] (6.0, 2.9) coordinate (s13) circle[];
			\fill[sample] (6.5, 2.6) coordinate (s14) circle[];
			\fill[sample] (7, 2.1) coordinate (s15) circle[];
			\fill[sample] (7.5, 1.8) coordinate (s16) circle[];
		\end{tikzpicture}
		\caption{Discrete samples}
		\label{example:discrete}

	\end{subfigure}
	\hfill{}
	\begin{subfigure}[c]{.48\textwidth}
		\centering
		\begin{tikzpicture}[shorten >=1pt, scale=.6, yscale=.6, xscale=0.7, auto]

			\draw[->] (0, 0) --++ (0, 5.0) node[anchor=north east]{$x$};
			\draw[->] (0, 0) --++ (8.0, 0) node[anchor=north]{$t$};
			\draw[dashed, color=red, semithick] (0, 4) --++ (8.0, 0);

			\fill[sample] (0, 1.8) coordinate (s1) circle[];
			\fill[sample] (0.5, 1.4) coordinate (s2) circle[];
			\fill[sample] (1, 1.9) coordinate (s3) circle[];
			\fill[sample] (1.5, 2.1) coordinate (s4) circle[];
			\fill[sample] (2, 2) coordinate (s5) circle[];
			\fill[sample] (2.5, 1.9) coordinate (s6) circle[];
			\fill[sample] (3, 1.4) coordinate (s7) circle[];
			\fill[sample] (3.5, 1.2) coordinate (s8) circle[];
			\fill[sample] (4, 0.8) coordinate (s9) circle[];
			\fill[sample] (4.5, 1.6) coordinate (s10) circle[];
			\fill[sample] (5, 2.5) coordinate (s11) circle[];
			\fill[sample] (5.5, 2.8) coordinate (s12) circle[];
			\fill[sample] (6.0, 2.9) coordinate (s13) circle[];
			\fill[sample] (6.5, 2.6) coordinate (s14) circle[];
			\fill[sample] (7, 2.1) coordinate (s15) circle[];
			\fill[sample] (7.5, 1.8) coordinate (s16) circle[];
			
			\node[nodraw] (s5') at (2.25, 4.8) {};
			
			\draw[signal] (s1) -- (s2);
			\draw[signal] (s2) -- (s3);
			\draw[signal] (s3) -- (s4);
			\draw[signal] (s4) -- (s5);
			\draw[signal] (s5) -- (s5');
			\draw[signal] (s5') -- (s6);
			\draw[signal] (s6) -- (s7);
			\draw[signal] (s7) -- (s8);
			\draw[signal] (s8) -- (s9);
			\draw[signal] (s9) -- (s10);
			\draw[signal] (s10) -- (s11);
			\draw[signal] (s11) -- (s12);
			\draw[signal] (s12) -- (s13);
			\draw[signal] (s13) -- (s14);
			\draw[signal] (s14) -- (s15);
			\draw[signal] (s15) -- (s16);

		\end{tikzpicture}

		\caption{Continuous behavior}
		\label{example:continuous}
	\end{subfigure}
	
	}
	\caption{Incompleteness}
\end{figure}
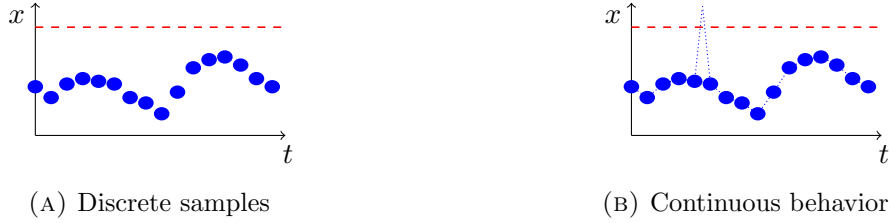

It would also be interesting to use refinement approaches such as CEGAR~\cite{CGJLV00} (Counterexample-Guided Abstraction Refinement) to refine both the time step and the bounding model.
\label{newtext:CEGAR}

\paragraph{Uncertain timestamps for online monitoring}
In contrast to our offline algorithm, our online algorithm monitoring assumes \emph{exact} timestamps:
this is not always realistic in all applications.
For example, triggering a sample via a shared network, or a long-distance communication (\eg{} with a satellite), can take a non-0 time, and result in a sample known with some uncertainty over the timestamp.
In that case, a future work is to not wait for the ``last second'' before triggering a new sample (as in \cref{algo:online}) but rather trigger a sample $\Delta$ time units prior to a possible safety violation, where $\Delta$ is some upper bound on the return delay between the monitor and the actual system under monitoring.

\paragraph{Discrete time vs.\ continuous time}
Another future work consists in increasing our guarantees, notably due to the \emph{continuous} nature of cyber-physical systems under monitoring.
Indeed, even with a rather fine-grained sampling showing no specification violation (\eg{} in \cref{example:discrete}), it can always happen that the actual \emph{continuous} behavior violated the specification (\eg{} in \cref{example:continuous}).
While setting discrete time steps at a sufficiently fine-grained scale will help to increase the confidence in the results of our approach, no absolutely formal guarantee can be derived.
Therefore, one of our future works is to propose some additional conditions for extrapolating (continuous) behaviors between consecutive discrete samples.
Also, improving the scope of our guarantees (in the line of, \eg{} \cite{DFS21}) is on our agenda.

Finally, in~\cite{WAH22TCPS}, the bounding model is given using linear hybrid automata, a formalism with a much more restricted dynamics than our approach, but featuring \emph{modes}, \ie{} changes of dynamics guarded by some constraints over the variables---which is not considered in our approach.
Extending our approach with modes (as in~\cite{WAH22TCPS}) is on our agenda, yielding a very expressive bounding model with dynamics beyond linear dynamics, \emph{and} modes.
However, this poses some technical difficulties, as the intersection of a set of behaviors with a guard (necessary to check a change of mode)
	is not proposed by the method from~\cite{ghosh1}.
A future work will be to envision over-approximated intersections.

\section*{Acknowledgment}
\noindent Bineet Ghosh was supported by the National Science Foundation (NSF) of the United States of America under grant number 2038960.
Étienne André is partially supported by the ANR-NRF French-Singaporean research program ProMiS (ANR-19-CE25-0015 / 2019 ANR NRF 0092)
and
ANR BisoUS (ANR-22-CE48-0012).

	\newcommand{\CCIS}{Communications in Computer and Information Science}
	\newcommand{\ENTCS}{Electronic Notes in Theoretical Computer Science}
	\newcommand{\FAC}{Formal Aspects of Computing}
	\newcommand{\FundInf}{Fundamenta Informaticae}
	\newcommand{\FMSD}{Formal Methods in System Design}
	\newcommand{\IJFCS}{International Journal of Foundations of Computer Science}
	\newcommand{\IJSSE}{International Journal of Secure Software Engineering}
	\newcommand{\IPL}{Information Processing Letters}
	\newcommand{\JAIR}{Journal of Artificial Intelligence Research}
	\newcommand{\JLAP}{Journal of Logic and Algebraic Programming}
	\newcommand{\JLAMP}{Journal of Logical and Algebraic Methods in Programming} %
	\newcommand{\JLC}{Journal of Logic and Computation}
	\newcommand{\LMCS}{Logical Methods in Computer Science}
	\newcommand{\LNCS}{Lecture Notes in Computer Science}
	\newcommand{\RESS}{Reliability Engineering \& System Safety}
	\newcommand{\RTS}{Real-Time Systems}
	\newcommand{\SCP}{Science of Computer Programming}
	\newcommand{\SOSYM}{Software and Systems Modeling ({SoSyM})}
	\newcommand{\STTT}{International Journal on Software Tools for Technology Transfer}
	\newcommand{\TCS}{Theoretical Computer Science}
	\newcommand{\TOPLAS}{{ACM} Transactions on Programming Languages and Systems ({ToPLAS})}
	\newcommand{\ToPNoC}{Transactions on {P}etri Nets and Other Models of Concurrency}
	\newcommand{\TOSEM}{{ACM} Transactions on Software Engineering and Methodology ({ToSEM})}
	\newcommand{\TSE}{{IEEE} Transactions on Software Engineering}

    \bibliographystyle{alphaurl}
	\bibliography{bibliofull}

\end{document}